\documentclass[12pt]{iopart}
\usepackage[T1]{fontenc}
\usepackage{bm}
\usepackage{svg}
\usepackage{float}
\expandafter\let\csname equation*\endcsname\relax
\expandafter\let\csname endequation*\endcsname\relax
\usepackage{amsmath}
\usepackage{graphicx}
\usepackage{dcolumn}
\usepackage{cite}
\usepackage{multirow}
\usepackage{footmisc}
\usepackage{amssymb}

\usepackage[dvipsnames]{xcolor}
\usepackage{hyperref}
\hypersetup{colorlinks=true, allcolors=RoyalBlue}
\usepackage{subcaption}
\usepackage{amsfonts}
\usepackage[capitalize]{cleveref}
\usepackage{physics}

\newcommand{\n}{\nonumber}
\usepackage{etoolbox}

\makeatletter
\newcommand{\fixappendix}{%
	\patchcmd{\l@section}{1.5em}{7em}{}{}%
	\patchcmd{\l@subsection}{2.3em}{7em}{}{}%
}
\makeatother

\newcommand{\thetitle}{Extreme value statistics and some applications in statistical physics}

\begin{document}
\title[\thetitle]{\thetitle}
\author{Marcin~Piotr Pruszczyk$^{1, 2}$ and Gregory Schehr$^{3}$}
\address{$^1$ SISSA --- International School for Advanced Studies, via Bonomea 265, 34136 Trieste, Italy}
\address{$^2$ INFN, via Bonomea 265, 34136 Trieste, Italy}

\address{$^3$ Sorbonne Universit\'e, Laboratoire de Physique Th\'eorique et Hautes Energies, 4 Place Jussieu, 75252 Paris Cedex 05, France}
\ead{mpruszcz@sissa.it, schehr@lpthe.jussieu.fr}
\begin{abstract} These notes are based on lectures delivered by G. Schehr at the XVIth School on Fundamental Problems in Statistical Physics (FPSP), held in Oropa (Italy) from 30 June to 11 July 2025. After a brief introduction to extreme value statistics (EVS) for independent and identically distributed (IID) random variables, we discuss several paradigmatic examples of strongly correlated systems where classical extreme value theory no longer applies. In particular, we focus on time series generated by random walks and Brownian motion, as well as on eigenvalue statistics in random matrix theory. Emphasis is placed on applications of EVS to fundamental problems in statistical physics and disordered systems, including the Random Energy Model, stochastic search problems, as well as fluctuating interfaces, and directed polymers in random media within the Kardar–Parisi–Zhang universality class.  
\end{abstract}

\tableofcontents
\markboth{\thetitle}{\thetitle}

\section{Introduction}
\label{sec:intro}

Rare events are uncommon by nature, but their consequences can be disproportionately large and a single extreme occurrence may dominate the overall outcome of a process. This is clearly illustrated in the {environmental sciences \cite{Katz_2002}}, where events such as earthquakes, heat waves, or unusually large ocean waves can cause significant damage despite their infrequent occurrence.
Similar issues arise in finance and actuarial science \cite{Novak_2011}, where one is often interested in identifying exceptionally high or low stock prices and understanding their impact.

Beyond these applied fields, rare events also play an essential role in the physics of complex systems \cite{Derrida_1981_REM,bouchaud1997universality,carpentier2001glass,majumdar2002extreme,le2003exact}; for recent reviews, see \cite{Majumdar_2020_EVS_review,amir2020thinking}. Many physical systems, especially those containing impurities or disorder, can be described in terms of an energy landscape with a large number of accessible states. At low temperatures, the system preferentially occupies the lowest-energy states, while higher-energy states contribute very little. As a result, the thermodynamic properties are largely determined by a small number of particularly low-energy configurations.

The dynamics of such systems is influenced in a similar way. When a system evolves in time, it must overcome energy barriers to move from one state to another. At low temperatures and over long time scales, the dynamics is controlled by the highest barriers. Consequently, physical observables such as the free energy or the relaxation time are often governed by extreme values rather than by typical ones. This naturally leads to the expectation that sample-to-sample fluctuations of these physical observables will carry clear signatures of extreme value statistics.

\begin{figure}
    \centering
    \includegraphics[width=0.7\linewidth]{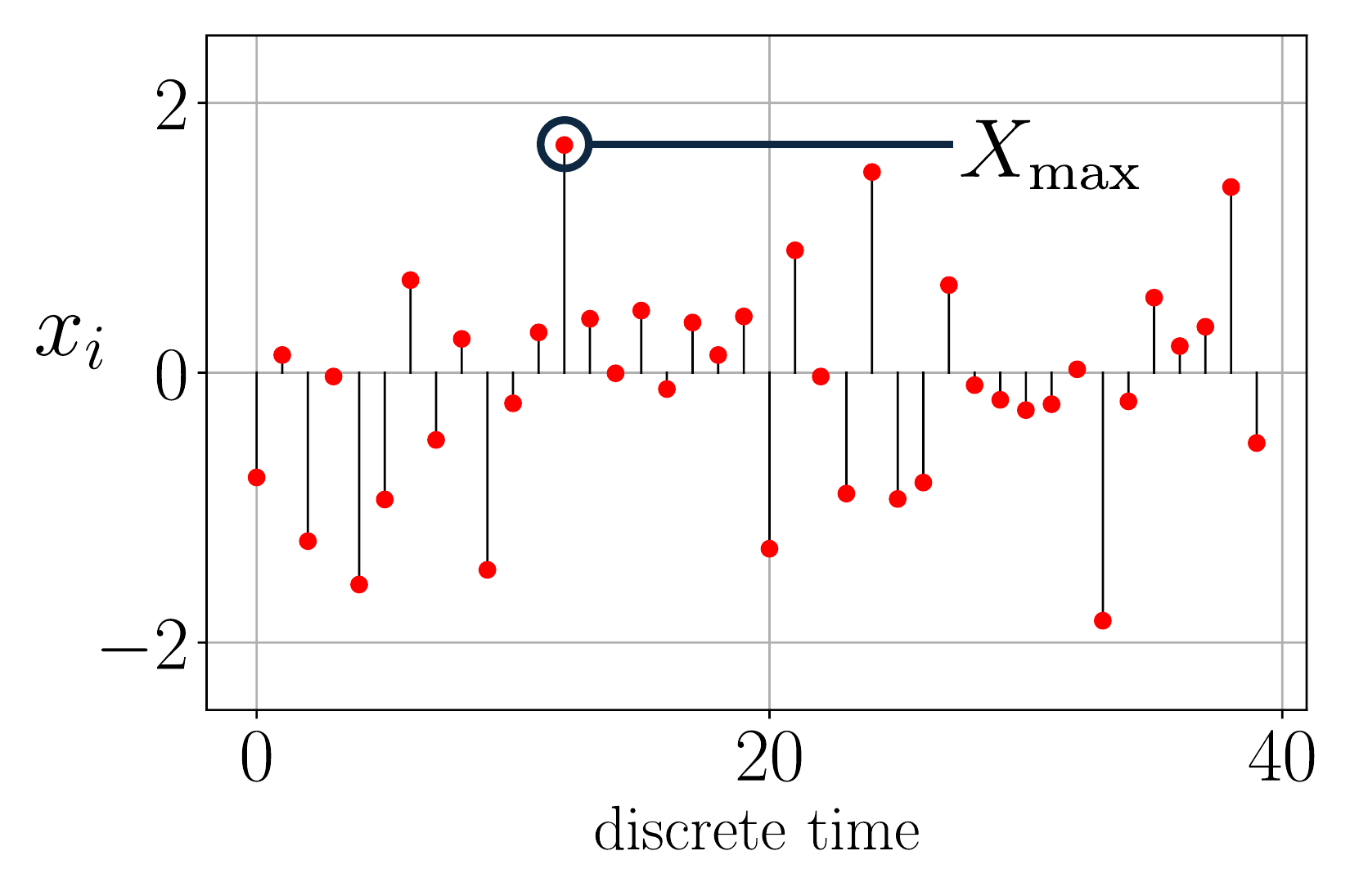}
    \caption{An example of a time series, here, a set of IID variables with Gaussian distribution with zero mean and unit variance.}
    \label{fig:iid_time_series}
\end{figure}

To illustrate the basic ideas, we consider a collection of $N$ random variables $x_1, x_2, \ldots, x_N$, 
{which may be interpreted} as successive observations of a stochastic process. We define the \emph{maximum} of this collection as
\begin{equation}
    X_{\mathrm{max}} = \max\{x_1, x_2, \ldots, x_N\},
    \label{eq:max_def}
\end{equation}
and the \emph{minimum} as
\begin{equation}
    X_{\mathrm{min}} = \min\{x_1, x_2, \ldots, x_N\}.
    \label{eq:min_def}
\end{equation}
Throughout this work, we interpret the index \(i\) as a discrete time variable, so that the ordered
set \(\{x_1, x_2, \ldots, x_N\}\) will be referred to as a \emph{time series}. An example of a time series with the resulting $X_{\max}$ is presented in \cref{fig:iid_time_series}. Such a framework arises naturally in many applications. For instance, \(x_i\) may denote the daily
maximum temperature recorded at a given location; in this case, \(X_{\mathrm{max}}\) represents
the highest temperature observed over a period of \(N\) days. In a financial context, \(x_i\) may
correspond to the value of a stock  
at the close of trading on day \(i\), so that
\(X_{\mathrm{max}}\) and \(X_{\mathrm{min}}\) describe the largest and smallest prices attained
during the observation window. As another example, in physics or engineering applications,
\(x_i\) may represent successive measurements of stress, energy, or signal amplitude, where the
extreme values characterize rare but potentially critical events.

The random variables \(X_{\mathrm{max}}\) and \(X_{\mathrm{min}}\) are referred to as
\emph{extreme values}. The study of their statistical properties forms the basis of extreme value statistics (EVS), see, e.g.,  Refs.~\cite{Majumdar_2020_EVS_review,amir2020thinking, Majumdar_Schehr_book, Gnedenko_1943, Gumbel_Book_1958, Resnick_book_1987, Leadbetter_book}. 
We begin by introducing the cumulative distribution function (CDF) of $X_{\rm max}$, defined as  
\begin{equation}
    Q_{\mathrm{max}}(w, N) = \mathrm{Prob.}(X_{\mathrm{max}} \leq w).
    \label{eq:CDF_def}
\end{equation}
In the case of  independent and identically distributed (IID) random variables $x_1, x_2, \ldots, x_N$, in the limit of $N \to \infty$,  appropriately rescaling $Q_{\max}$ with $N$ renders its limiting form that falls into one of three universality classes, namely the Gumbel, the Weibull, or the Fr\'echet class, which we discuss in Section~\ref{subsec:universality_classes} below. This fact, known as Gnedenko’s classical law of extremes \cite{Gnedenko_1943}, can be regarded as the EVS counterpart of the well-known Central Limit Theorem (CLT) describing the limiting distribution of a sample mean of IID variables. For a detailed discussion of the connections between these two types of observables, we refer the reader to Ref. \cite{bertin2006generalized}.
  
The description of the statistical properties of \(X_{\max}\) in terms of the three classical
universality classes of the extreme value theory, although elegant, is not exhaustive. In many
applications, one is instead faced with \emph{strongly correlated} random variables, for which
these results no longer apply. As an illustrative example from actuarial science, first consider a simple model for the price of a
stock, described by the stochastic process \cite{Novak_2011, Samuelson_1965, Black_1973}
\begin{equation}
    S_n = \exp(\sigma x_n),
\end{equation}
where \(\sigma > 0\) is the volatility, and \(x_n\)  evolves according to the recursive relation
\begin{equation}
    x_n = x_{n-1} + \eta_n,
    \label{eq:time_ser}
\end{equation}
with \(\{\eta_n\}\) being a sequence of independent Gaussian random variables with zero mean and the correlation function
\(\langle \eta_n \eta_{n'} \rangle = \delta_{n,n'}\), where \(\delta_{n,n'}\) denotes the
Kronecker delta. The process \(\{x_n\}\) is therefore a discrete-time random walk for which
\(\langle x_n x_{n'} \rangle = \min(n,n')\). As a consequence, the stock prices \(S_n\) and
\(S_{n'}\) are strongly correlated for \(n \neq n'\). Owing to these correlations, one expects
—and we shall show below—that the extreme value statistics of \(S_n\) (or equivalently of
\(x_n\)) are not described by any of the classical laws derived for IID random variables. In Section \ref{sec_RW}, we will discuss in more detail the EVS of time series generated by a random walk such as $x_n$ in Eq.~(\ref{eq:time_ser}).

\begin{figure}[t]
    \centering
    \begin{tabular}{cc}{{\includegraphics[width=0.5\linewidth]{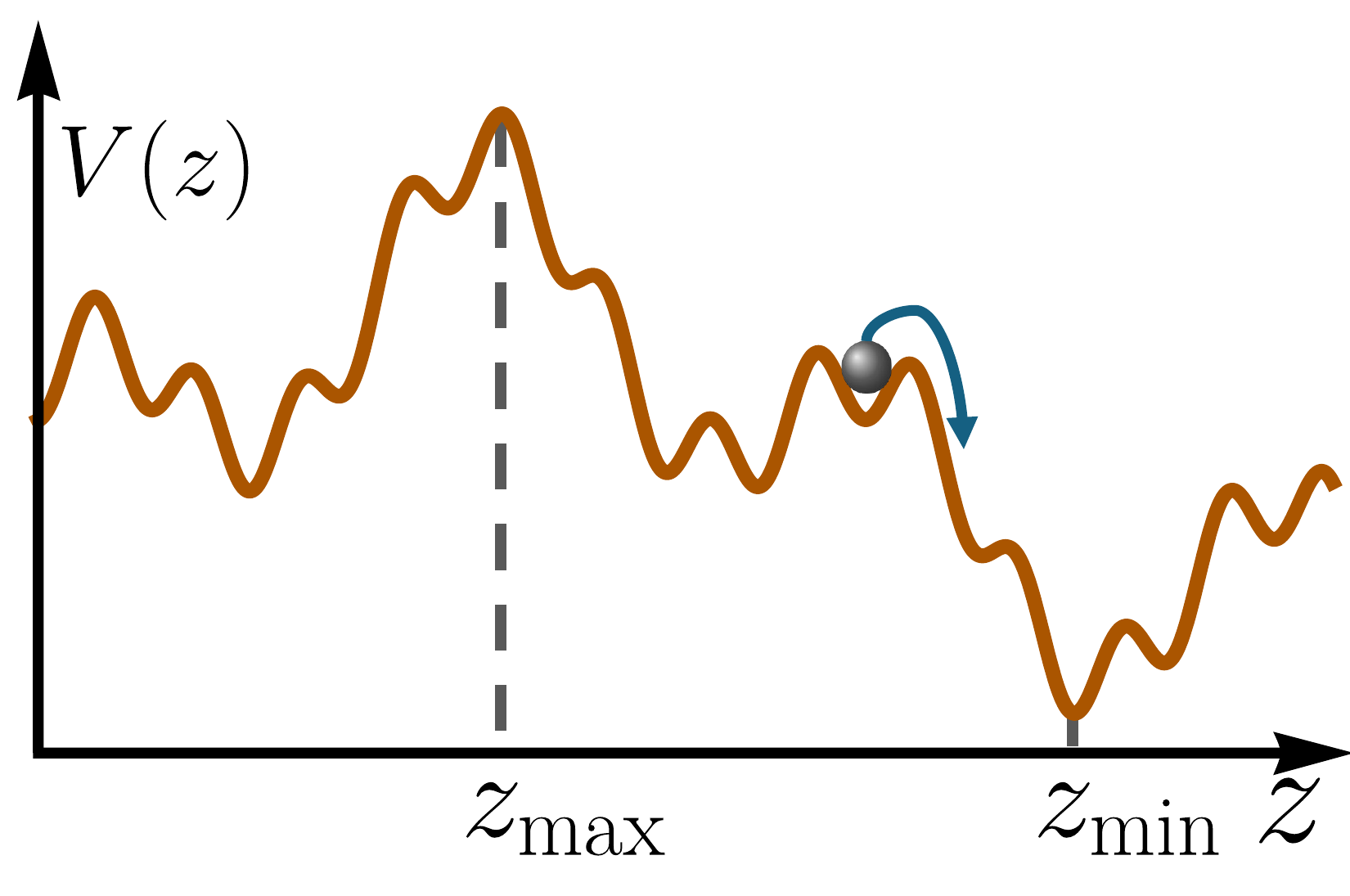}}}&\raisebox{0.2in}{{\includegraphics[width=0.45\linewidth]{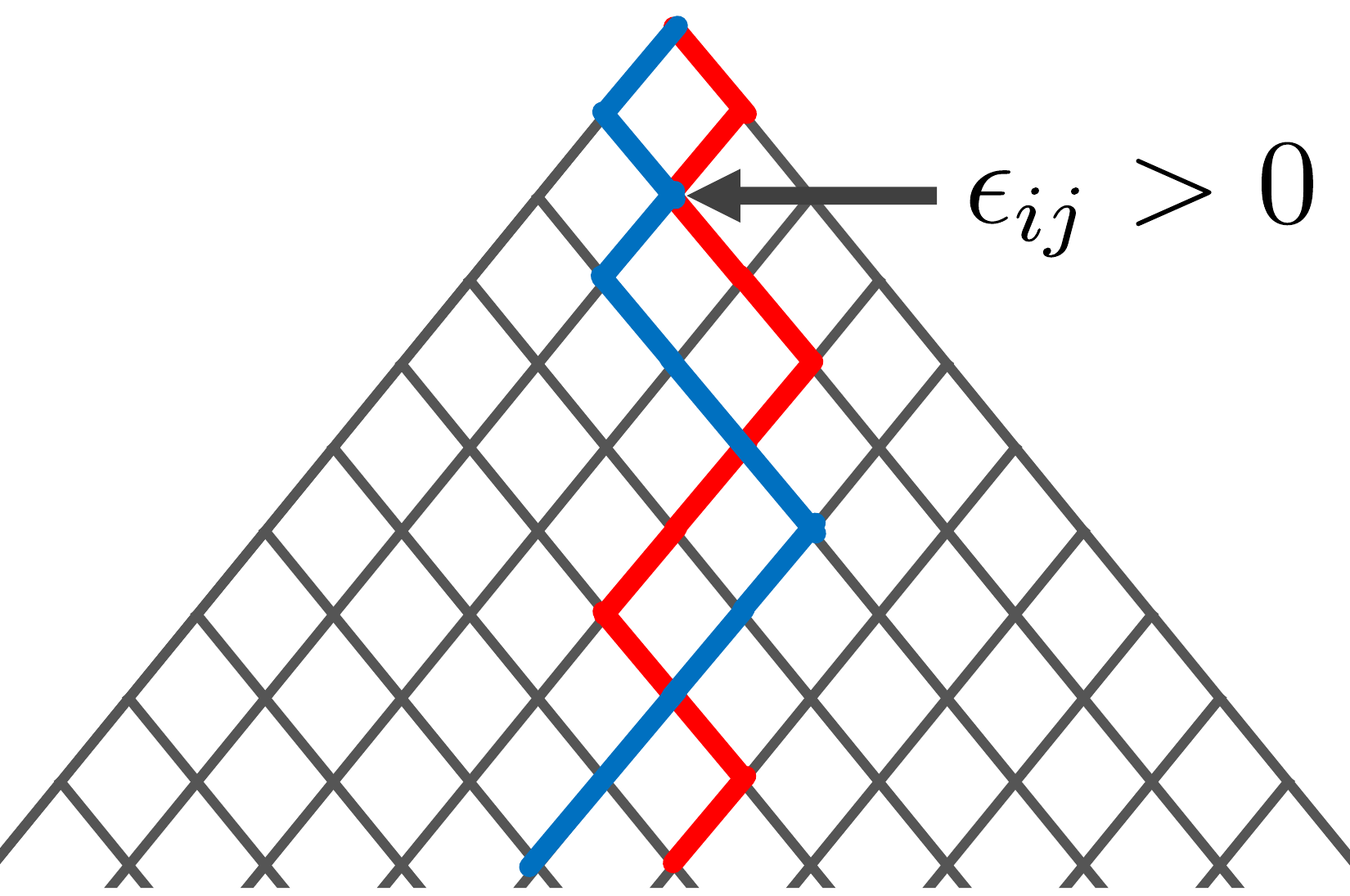}}}
        \\[-2mm]
         (a) & (b) 
    \end{tabular}
\caption{Examples of  disordered systems where the description of the extreme value statistics plays a key role in understanding their physical properties.  In panel (a), a cartoon of a particle diffusion in a rough energy landscape (e.g., {Sinai's} model). Here, $z_{\max}$ and $z_{\min}$ denote respectively the positions of the maximum and the minimum of the potential $V(z)$.   In panel (b), a representation of two configurations of a directed polymer on a square lattice. Note that they are correlated, since they share sites after the second, fifth, and eighth steps.}
\label{fig:disordered}
\end{figure}

In statistical physics, particularly in the context 
of disordered systems, such as spin glasses, particles diffusing in random potentials, or polymers in random media, 
one usually has to deal with strongly correlated random variables. A well-known model of a particle's motion in an energy landscape,   Sinai's model \cite{Sinai_1982}, may serve as an example. There, the particle's position $z(t)$ evolves according to overdamped dynamics in a random potential $V(z)$, which, as a function of the position $z$,  
is a Brownian motion -- the continuum limit of the random walk described above in Eq. (\ref{eq:time_ser}). Thus, the energy landscape $V(z)$ exhibits strong spatial correlations. At low temperatures, the particle will be trapped in the vicinity of $z_{\min}$ -- see Fig. \ref{fig:disordered}(a). Hence, these extreme value statistics are reflected in the probability distribution of the particle's position in equilibrium. Similarly, the largest energy barrier of the landscape $V(z)$ --- see \cref{fig:disordered}(a)--- provides information about the transport properties of the particle in this disordered medium. 
As another emblematic example, let us consider  a two-dimensional directed polymer in a random medium \cite{Kardar_1987, Derrida_1988, Imbrie_1988, Cook_1989, Mezard_1990, Fisher_1991,Krug_1998,  Majumdar_Les_Houches_2007},  which is modeled by a two-dimensional lattice with random quenched IID energies $\epsilon_{i,j}$ on each site $(i,j)$. Two configurations of such polymers are presented in \cref{fig:disordered}(b). The energy of an $n$-step directed polymer on such a lattice is given  by the sum of energies of the sites visited by the polymer 
\begin{equation}
    E(\mathcal{C}) = \sum_{\langle i,j\rangle \in  \mathcal{C}} \epsilon_{i,j},
\end{equation}
which is a random variable. It turns out that the optimal (highest) energy of the polymer
\begin{equation}
    E_{\mathrm{max}} = \max_{\mathcal{C}}  E(\mathcal{C})
\end{equation}
follows the same statistics as the largest eigenvalue of a random matrix, which we will discuss below in more detail in Section~\ref{sec_RM}.

The two examples illustrated in Fig. \ref{fig:disordered}, together with the additional cases discussed in Section~\ref{sec_RW} and Section~\ref{sec_RM}, demonstrate that EVS emerge naturally across a wide range of fundamental problems in statistical physics. Conversely, as we show below, many standard questions in EVS can be formulated as statistical mechanics problems \cite{Majumdar_Schehr_book}.

\begin{figure}
    \centering
    \begin{tabular}{cc}
        \raisebox{0.3in}{\includegraphics[width=0.5\linewidth]{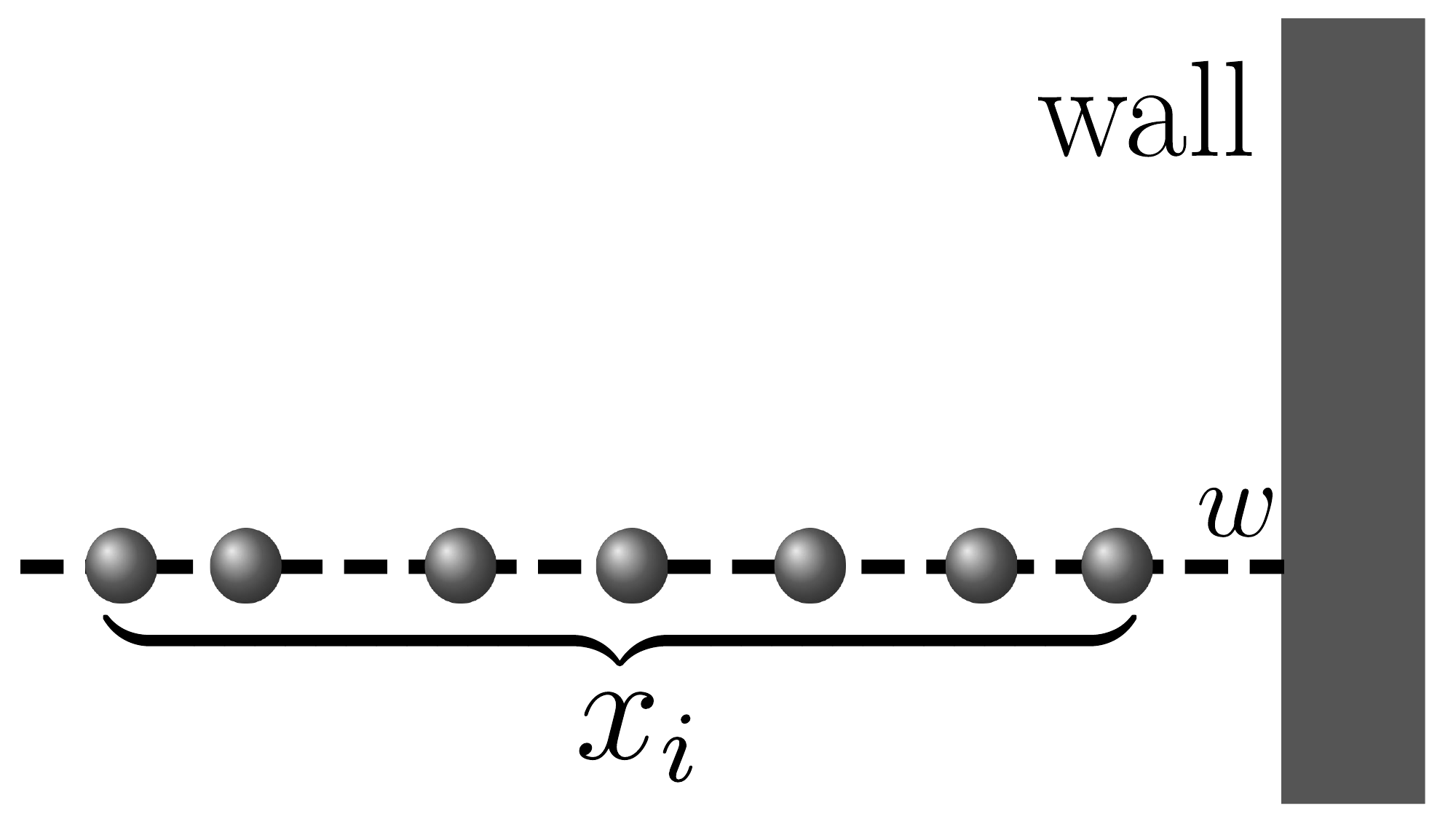}}&
        \includegraphics[width=0.5\linewidth]{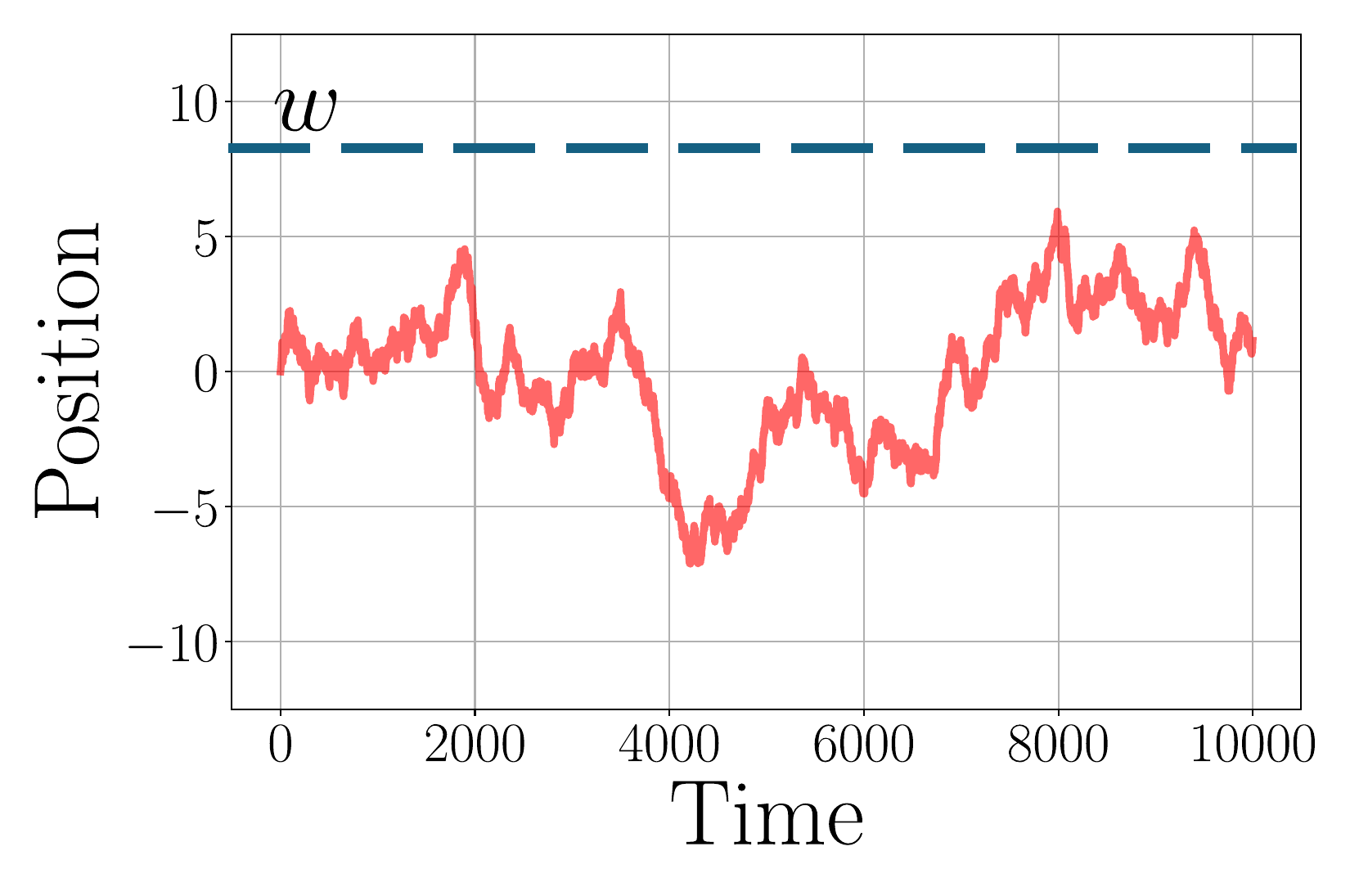}
        \\[-2mm]
         (a) & (b) 
    \end{tabular}
\caption{Two physical interpretations of the CDF of $X_{\max}$. In panel (a), we provide a cartoon of  $N$ particles on a line in the presence of the wall placed at $w$. Associating an energy to configurations of the particles' positions according to \cref{eq:E_from_Pjoint} renders the canonical partition function of the system equal to $Q_{\max}(w, N)$, as indicated in \cref{eq:CDF_as_partition}. 
In panel (b), we plot a trajectory of a Brownian motion. The survival probability of the trajectory staying below $w$ is equivalent to $Q_{\max}(w,N)$, with $N$ denoting the number of time steps. }
\label{fig:phys_interpret}
\end{figure}

\vspace*{0.5cm}
\noindent{\bf EVS as a basic statistical mechanics problem.}
Let us denote the joint probability distribution function (PDF) of $x_i$'s by $P_{\mathrm{joint}}(x_1, x_2, \ldots,  x_N)$. We notice that the CDF of $X_{\max}$
can be expressed as
\begin{align}
     Q_{\mathrm{max}}(w, N) &= \mathrm{Prob.}(x_1 \leq w, x_2 \leq w, \ldots, x_N \leq w) \n \\ &=   \int_{-\infty}^w \mathrm{d}x_1\int_{-\infty}^w \mathrm{d}x_2 \ldots \int_{-\infty}^w \mathrm{d}x_N P_{\mathrm{joint}}(x_1, x_2, \ldots,  x_N)\;.
     \label{eq:CDF_from_Pjoint}
\end{align}
This crucial expression (\ref{eq:CDF_from_Pjoint}) is the starting point of most of the analytical approaches in EVS.  It can be understood as follows.   
The event $\{X_{\max} \leq w \}$ is exactly the same as the event $\{x_1 \leq w, \; x_2 \leq w, \; \cdots, \; x_{N} \leq w \}$. This is because, if the maximum of a set is $ \leq w$, it necessarily implies that all elements of this set are $\leq w$. Similarly, if all elements of a set are $\leq w$, this necessarily means that their maximum $\leq w$. The mapping between these two events implies the relation in {Eq.~(\ref{eq:CDF_from_Pjoint})}. 
In order to make a connection with equilibrium statistical mechanics, we  introduce an `energy' $E$ of a configuration of $x_i$'s according to
\begin{equation}
    E(x_1, x_2, \ldots,  x_N) = - \ln P_{\mathrm{joint}}(x_1, x_2, \ldots,  x_N).
    \label{eq:E_from_Pjoint}
\end{equation}
This allows us to cast the expression for the cumulative distribution function \eqref{eq:CDF_from_Pjoint} in  the form 
\begin{equation}
    Q_{\mathrm{max}}(w, N ) =  \int_{-\infty}^w \mathrm{d}x_1\int_{-\infty}^w \mathrm{d}x_2 \ldots \int_{-\infty}^w \mathrm{d}x_N \,e^{-E(x_1, x_2, \ldots, x_N)} \;.
    \label{eq:CDF_as_partition}
\end{equation}
The integrand on the right-hand side of \cref{eq:CDF_as_partition} can  be interpreted as the Boltzmann weight of a configuration of a one-dimensional gas with the corresponding energy given by \cref{eq:E_from_Pjoint}. Notice that  the upper limit of integration in \cref{eq:CDF_as_partition} corresponds to an infinite potential wall placed at $w$, as we indicate in the cartoon of the resulting physical system in \cref{fig:phys_interpret}(a). 
 Hence, $Q_{\mathrm{max}}(w, N)$ can be obtained by evaluating the partition function of the corresponding one-dimensional gas in the presence of a hard wall. 

In a similar spirit, the cumulative distribution function of the maximum of a time series can be related to the theory of first-passage time of stochastic processes \cite{Majumdar_Schehr_book}. Indeed, $Q_{\mathrm{max}}(w,N)$ can be interpreted as the \textit{survival probability} of a stochastic process staying below $w$ after $N$ time steps, as presented in \cref{fig:phys_interpret}(b). This interpretation allows us to draw connections between EVS and the studies of persistence and first-passage problems, which we will discuss further in the context of random walks and Brownian motion in Section \ref{sec_RW}.

We end this section by noting that one can not only investigate the statistics of the maximal or minimal value of the time series, but also the statistics of the $k$-th largest value, which we denote as $M_{k,N}$, satisfying
\begin{align}
    X_{\mathrm{max}} = M_{1,N} > M_{2,N} > \ldots > M_{N,N} = X_{\mathrm{min}}  
\end{align}
This study of the so-called order statistics goes, however, beyond the scope of these lecture notes, and we refer the interested reader to Ref.~\cite{Arnold_book, Majumdar_Schehr_book, David_book,SM_ideas} for more details.

The rest of these notes is organized as follows. In \cref{sec:EVS_IID}, a brief introduction to EVS of a series of IID random variables is given. 
In \cref{sec_RW},  
the cases of random walks and Brownian motion are discussed, and their connection to first-passage problems and survival probabilities is explained. In \cref{sec_RM}, eigenvalue statistics of random matrices and Tracy-Widom laws are presented, along with their applications to physical problems. Finally, in \cref{sec:conclusions}, we present conclusions as well as more recent developments.

\section{EVS of time series with independent and identically distributed entries} 
\label{sec:EVS_IID}
\subsection{The typical value \texorpdfstring{$\mu_N$}{mu_N} }
\label{subsec:EVS_IID_typical}
\begin{figure}
    \centering
    \begin{tabular}{cc}
        {\includegraphics[width=0.47\linewidth]{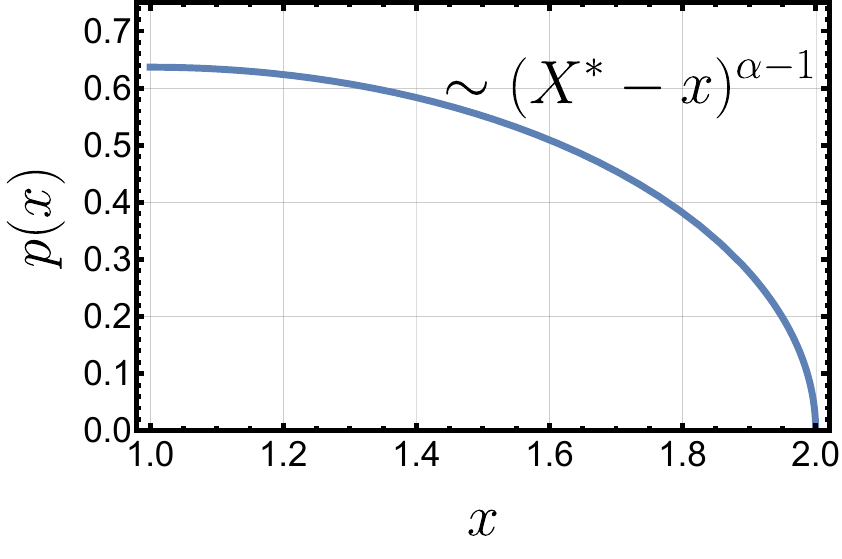}}&
        \raisebox{.2in}{\includegraphics[width=0.47\linewidth]{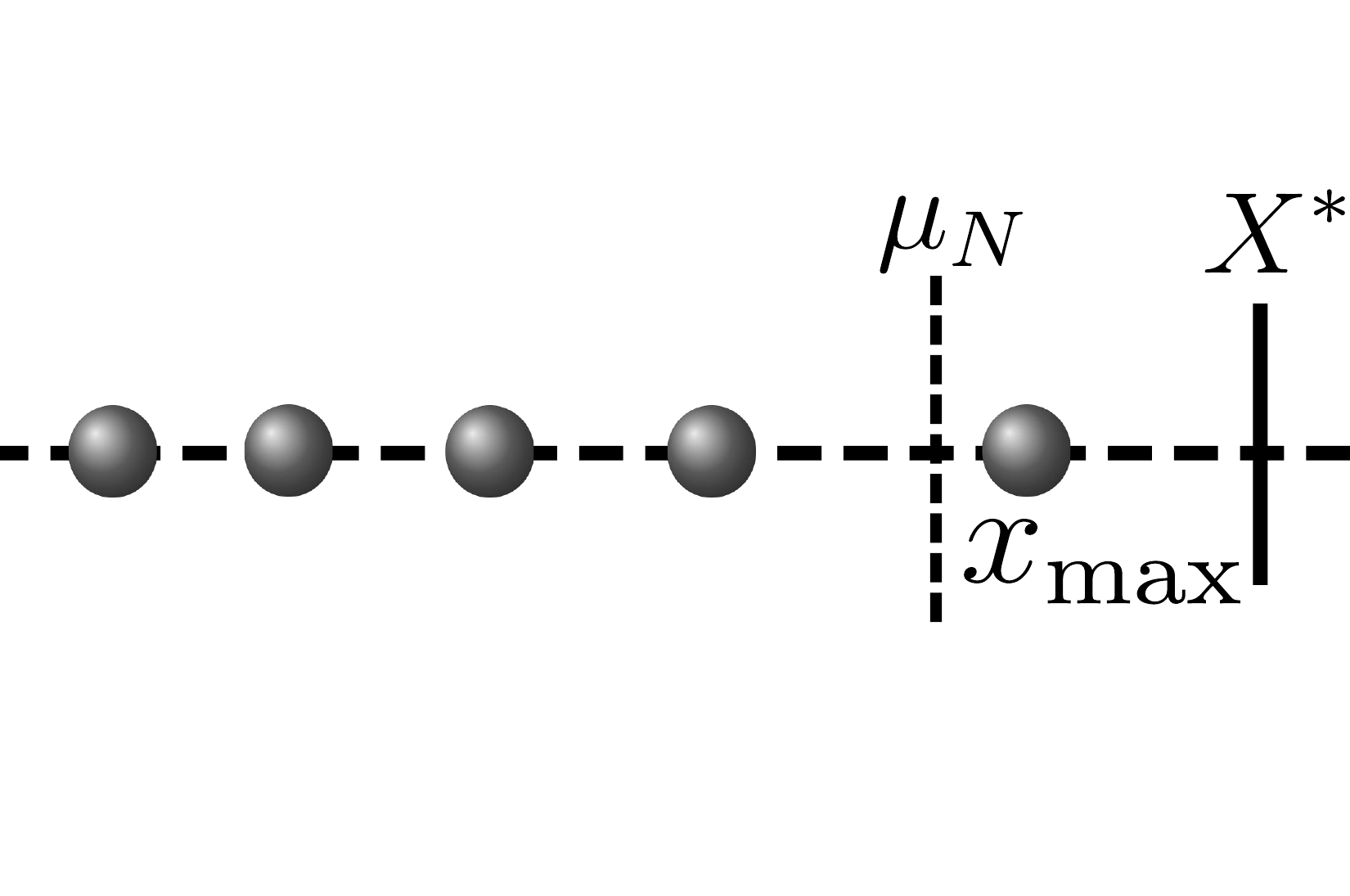}}
        \\[-2mm]
         (a) & (b) 
    \end{tabular}
\caption{In panel (a), we plot the edge of a distribution $p(x) = 2/\pi \sqrt{1 - (x-1)^2}$ which has a support bounded at $X^* = 2$, and which vanishes as $p(x)\propto (X^*-x)^{1/2}$ as $x \to X^{*-}$.   As we discuss further, such a parent distribution falls into the Weibull universality class with {$\alpha = 3/2$}, c.f.,  \cref{eq:Wiebull_parent}. In panel (b), we provide a cartoon illustrating  that  for $N$ large enough,  there is almost surely $\mathcal{O}(1)$ of variables $x_i$ in the interval $[\mu_N, X^*]$. }
\label{fig:typ_def}
\end{figure}

In the case of IID variables, it is convenient to consider the cumulative distribution function of a single variable
\begin{equation}
    {\rm Prob.}(x_i \leq x) = \int_{-\infty}^x \mathrm{d}x' p(x') \;,
\end{equation}
expressed in terms of the associated PDF $p(x)$, also called the parent distribution, which, for simplicity, we assume here to be continuous. Since the variables $x_i$'s are independent, their joint PDF factorizes
\begin{equation}
    P_{\mathrm{joint}}(x_1, x_2, \ldots, x_N) = p(x_1)p(x_2)\ldots p(x_N)\;.
    \label{eq:pdf_fact}
\end{equation}
Let us now suppose that the support of the parent distribution is bounded from above, i.e., that there exists $X^* \in \mathbb{R}$ such that $p(x) =0$ for $x>X^*$, and $\forall \epsilon>0$, $ p(X^* - \epsilon)>0$. In principle, the upper edge can be infinite, i.e., $X^* \to \infty$, but the discussion turns out to be simpler if one considers $X^*$ finite. An example of such a distribution with $X^*=2$ is presented in \cref{fig:typ_def}(a). It turns out that in such a case
\begin{equation}
    X_{\mathrm{max}}\xrightarrow[N \to \infty]{} X^* \;.
\end{equation}
When $N$ is large, $N\gg 1$, but finite, one can estimate how close $X_{\mathrm{max}}$ is to $X^*$ by introducing the typical value $\mu_N$ such that almost surely a number of $\mathcal{O}(1)$ of $x_i$'s are in the interval $[\mu_N, X^*]$, as shown in a cartoon provided  in \cref{fig:typ_def}(b). This yields the relation
\begin{equation}
N\int_{\mu_N}^{X^*}\mathrm{d}x\,p(x) \approx 1 \;
\label{eq:typ_val_def}
\end{equation}
which is particularly useful, since 
it allows us to estimate the typical scale of $X_{\max}$. Quite remarkably, while this relation (\ref{eq:typ_val_def}) holds a priori for IID random variables, it has been shown to hold in a much wider set of problems, including, in particular, EVS for strongly correlated systems \cite{Majumdar_2020_EVS_review,Majumdar_Schehr_book}.

Below, we provide expressions for the typical value for several parent distributions and we generalize to the case of $p(x)$ with an unbounded support ($X^* \to \infty$).   

\vspace*{0.5cm}\noindent{\bf Uniform distribution.} First, let us consider the uniform distribution 
\begin{equation}
    p(x) = \begin{cases}
        1, \quad   x \in [0,1], \\ 0, \quad \mbox{otherwise}, 
    \end{cases}
\end{equation}
for which $\mu_N = 1 - 1/N$, which suggests that the fluctuations of the maximal value $1-X_{\max}$ are of the order $\mathcal{O}(N^{-1})$. 

\vspace*{0.5cm}\noindent{\bf Exponential distribution.} 
In this case, the parent distribution has  a semi-infinite support over $[0,+\infty)$ -- hence $X^* = +\infty $ -- and it reads
\begin{equation} \label{eq:p_exp}
 p(x) = \begin{cases}
     e^{-x}, \quad \, \, \, \mbox{for} \quad x\geq 0\;, \\ 0 \;, \qquad  \mbox{for} \quad x<0 \;.
 \end{cases}
\end{equation}
Plugging this expression for $p(x)$ into Eq. (\ref{eq:typ_val_def}), one easily finds  $\mu_N = \ln N$, which thus grows unboundedly as $N \to \infty$. 

\vspace*{0.5cm}\noindent{\bf Gaussian distribution.}
In the case of the Gaussian distribution
\begin{equation} \label{eq:p_Gauss}
    p(x) = \frac{1}{\sqrt{2 \pi \sigma^2}}e^{-\frac{x^2}{2 \sigma^2}}, 
\end{equation}
which is defined on the full real line, i.e., with $X^* = +\infty$ also in this case, one finds from Eq. (\ref{eq:typ_val_def}) that the typical value diverges as $N \to \infty$ as $\mu_N = \sigma \sqrt{2 \ln N} + \mathcal{O}(\ln \ln N/\sqrt{\ln N})$. Note, however, that $\mu_N$ grows more slowly than in the exponential case. This is physically reasonable since the probability of observing large values for a Gaussian distribution is smaller than for an exponential one. We also remark that, in this case, finite-size corrections turn out to be particularly large \cite{Gyorgi_2008,gyorgyi2010renormalization}.

An interesting application of the EVS of IID Gaussian random variables arises in the context of the Random Energy Model (REM)~\cite{Derrida_1981_REM}, which is a well-known and useful toy model of a mean-field spin-glass. There, one considers a partition function of the form
\begin{equation} \label{eq:Z_REM}
    Z_{\rm REM}= \sum_{i=1}^{M}e^{-\beta E_i}, \quad \mbox{with} \quad M=2^N,
\end{equation} 
where $\beta = 1/k_BT$, $k_B$ is the Boltzmann constant, $T$ is the temperature, and $E_i$'s can be interpreted as the energies of configurations of $N$ Ising spins -- hence their number $M = 2^N$. We assume them to be IID Gaussian variables with a vanishing mean and variance which scales with the number of spins according to
\begin{equation} \label{eq:Eij}
    \langle E_i E_j \rangle = \delta_{i,j} \frac{N J^2}{2}\;.
\end{equation}
Here, the scaling with $N$ is chosen so that the model defined by (\ref{eq:Z_REM}) has a well-defined thermodynamic limit $N \to \infty$ (see below), and $J$ is the coupling constant that sets the energy scale of the spin-spin interactions. A central thermodynamic observable is the quenched free energy per unit spin
\begin{equation}
    f_q = -\lim_{N\to \infty} \frac{k_BT}{N} \overline{\ln Z}\; ,
    \label{eq:f_q}
\end{equation}
where $\overline{\cdots}$ denotes the average over the disorder, i.e., over the distribution of $E_i$'s, with correlations given as in (\ref{eq:Eij}). 
The resulting expression for $f_q$ turns out to be of the form~\cite{Derrida_1981_REM}
\begin{equation} \label{eq:fq}
    f_q = \begin{cases}
        -k_BT \ln 2 - \dfrac{J^2}{4 k_B T}, \quad \, \mbox{for } \quad T>T_c = \frac{J}{k_B}\frac{1}{2 \sqrt{\ln 2}}\;, \\ -J\sqrt{\ln 2} \;, \quad \quad \quad\quad\quad \;  \mbox{for } \quad T<T_c \;.
    \end{cases}
\end{equation}
While the finite temperature analysis of (\ref{eq:f_q}) requires a careful computation -- which can be carried out using different methods (see, for instance, \cite{parisi2004course}) -- the zero temperature result, i.e., the second line of \cref{eq:fq}, can be obtained rather simply using EVS for IID Gaussian random variables as follows. 
As $T \to 0$, the free energy of the system converges to the ground state energy of the system, and one expects
\begin{eqnarray} 
\lim_{T \to 0} f_q = \frac{\overline{E_{\min}}}{N} \;,
\end{eqnarray}
where $E_{\min} = \min_{1 \leq i \leq M} E_i$. We recall that, here, the random energies $E_i$'s are IID Gaussian random variables with zero mean and variance $\sigma_{\rm REM}^2 = N J^2/2$, which can be read off from Eq.~(\ref{eq:Eij}). Using the symmetry of the Gaussian distribution under the {sign change} $E \to -E$, which implies that $E_{\min}$ has the same statistics as $-E_{\max}$, one finds, using the aforementioned result for EVS of Gaussian random variables derived from~(\ref{eq:typ_val_def}), that
\begin{eqnarray} \label{eq:fq_REM}
\lim_{T \to 0} f_q = \frac{\overline{E_{\min}}}{N} \approx - \frac{\mu_M}{N} \approx - \frac{\sigma_{\rm REM} \sqrt{2 \ln M}}{N} = - J \sqrt{\ln 2} \;,
\end{eqnarray}
which indeed reproduces the result given in the second line of Eq. (\ref{eq:fq}). 

\vspace*{0.5cm}\noindent{\bf Cauchy distribution.}
Another interesting example is the Cauchy distribution
\begin{equation}
    p(x) = \frac{1}{\pi}\frac{1}{1+x^2} \;.
\end{equation}
In this case, the typical value of the maximum $\mu_N$ can be easily obtained from Eq.~(\ref{eq:typ_val_def}), with, again, $X^* = + \infty$, and it 
simply yields $\mu_N = N/\pi  + \mathcal{O}(1/N)$.

\subsection{An exact formula for the CDF}
In the case of IID variables, the exact formula for $Q_{\max}(w, N)$ in \cref{eq:CDF_from_Pjoint}
can actually be written rather explicitly. Indeed, using the factorization form of the joint distribution as given in \cref{eq:pdf_fact}, a straightforward calculation leads to 
\begin{align}
   Q_{\mathrm{max}}(w, N) &= \int_{-\infty}^w \mathrm{d}x_1\int_{-\infty}^w \mathrm{d}x_2 \ldots \int_{-\infty}^w \mathrm{d}x_N \, P_{\mathrm{joint}}(x_1, x_2, \ldots,  x_N)  \\
    &= \prod_{i = 1}^N{\rm Prob.}(x_i<w) = \left[\int_{-\infty}^w \mathrm{d}x \, p(x) \right]^N.
     \label{eq:CDF_from_Pjoint_1}
\end{align}
The expression of the CDF above allows us to express the PDF of $X_{\max}$ in a neat form
\begin{align}
   P_{\mathrm{max}}(w, N) &= \frac{\rm d}{\mathrm{d}w}Q_{\max}(w,N) = Np(w)\left[\int_{-\infty}^w \mathrm{d}x \, p(x) \right]^{N-1}
     \label{eq:PDF_from_Pjoint_1}. 
\end{align}

To provide a concrete  example, let us consider  the case where the parent distribution is a pure exponential, as in \cref{eq:p_exp}, for which we have seen above that the typical value of $X_{\max}$ is $\mu_N = \ln N$. Making use of the result in \cref{eq:CDF_from_Pjoint_1}, we write $Q_{\max}$ in the form 
\begin{equation}
    Q_{\max}(w,N) = \left[\int_{0}^w \mathrm{d}x \, e^{-x} \right]^N = \left[1-e^{-w} \right]^N.
\end{equation}
To investigate the fluctuations around the typical value, we consider the CDF as a function of $w = \mu_N + z = \ln N +z$, which yields
\begin{equation}
    Q_{\rm max}\left(w = \ln N + z\right) = \left[1-\frac{1}{N}e^{-z} \right]^N \xrightarrow[N\to \infty]{}e^{-e^{-z}}. 
\end{equation}
This is the celebrated Gumbel distribution \cite{Gumbel_Book_1958}, which we plot in \cref{fig:EVS_class_example}. 
We note that, importantly,  while considering the $N \gg 1$ regime, 
we expressed the CDF in terms of  a scaling variable of the form
\begin{equation}
    y = \frac{X_{\mathrm{max}}- a_N}{b_N} \label{eq:scaled_var}
\end{equation}
for which the CDF admits an $N$-independent form as $N \to \infty$. Notice that without such a rescaling, the right-hand side of \cref{eq:CDF_from_Pjoint_1} converges  trivially to $0$  in the limit of $N \to \infty$, since it is a positive number smaller than $1$ to the $N$-th power. In the case of the exponential distribution considered above, we found $a_N = \ln N$ and $b_N = 1$. 
The goal of the EVS for IID random variables, as discussed in the following subsection, is precisely to find the scaling coefficients $a_N$ and $b_N$, and the limiting distribution $G(y)$ such that $\lim_{N \to \infty} Q_{\max}(a_N  + b_Ny,N) = G(y)$ for any given parent distribution $p(x)$.
\subsection{Universality classes for IID variables}
\label{subsec:universality_classes}
It turns out that, for IID variables, there are only three such limiting distributions 
\begin{equation}
    G(y) = G_{\rho}(y)\;, \quad {\rm with}  \qquad \rho = \mathrm{I}, \mathrm{II}, \mathrm{III},
\end{equation}
with the corresponding PDFs denoted by $g_{\rho}(y) = G_{\rho}'(y)$. The CDF of the maximum, after appropriate rescaling, falls into one of the three universality classes depending on the large-$x$ behavior of the parent distribution. 

\vspace*{0.5cm}
\noindent{\bf The Gumbel universality class I.} First, let us consider the case of  $p(x)$ with an  unbounded support,  which satisfies
\begin{equation}
    p(x) \ll x^{-\eta}\;, \quad  \forall  \eta>0 \quad \mbox{as} \quad  x \to \infty\;,
\end{equation}
i.e., it decays, for large $x$, faster than any power-law. The exponential and Gaussian distributions studied above naturally fall into this class. This is the so-called Gumbel universality class for which the limiting distribution is a Gumbel law \cite{Fisher_1928, Gumbel_Book_1958}, namely,
\begin{equation}
    G_{\mathrm{I}}(y) = e^{-e^{-y}}\;, \qquad g_{\mathrm{I}}(y) = e^{-y -e^{-y}}\;.
    \label{eq:gumbel_dist}
\end{equation}
The centering and scaling coefficients $a_N$ and $b_N$, yielding a nontrivial limiting CDF as $N \to \infty$, are given by 
\begin{align}    
    &a_N = \mu_N \;, \\ 
    &b_N = \frac{\left(\int_{a_N}^{\infty}\mathrm{d}x (x - a_N)p(x)\right)}{\left(\int_{a_N}^{\infty}\mathrm{d}x \, p(x)\right)}\;,
    \label{eq:a_n_gumbel}
\end{align}
where $\mu_N$ is the typical value introduced in  \cref{eq:typ_val_def}.
For exponentially distributed random variables, as in (\ref{eq:p_exp}), for which $a_N = \mu_N = \ln N$, one easily checks from \cref{eq:a_n_gumbel} that $b_N = 1$, as anticipated above. Hence, in this case, the fluctuations of the maximum around its typical value $a_N = \mu_N$ remain sizeable and of order $O(1)$ as $N \to \infty$. On the other hand, for centered Gaussian random variables of variance $\sigma^2$, as in Eq.~(\ref{eq:p_Gauss}), for which the typical value is $a_N = \mu_N \approx \sigma\sqrt{2 \ln N}$, one finds from Eq.~(\ref{eq:a_n_gumbel}) that $b_N \approx 1/(\sigma \sqrt{2 \ln N})$ as $N \to \infty$. Therefore, in contrast with the exponential case, the fluctuations of $X_{\max}$ around its typical value $a_N = \mu_N$ vanish as $N \to \infty$, and in many cases the distribution of $X_{\max}$ can be well approximated by a Dirac delta function centered at $a_N$. 

An interesting application of this property of `concentration around the mean' for the EVS of Gaussian IID variables occurs in the simple predator-prey model where a `lamb' is chased by a `pride of lions', as studied, for instance, in \cite{Krapivsky_1996_lamb_lion, Redner_1999_lamb_lion}. There, one considers $N$ Brownian particles in one dimension (lions) with positions $x_i(t)$, each starting at the origin and diffusing independently with diffusion constant $D$.
Additionally, at some $x_0>0$, there is a single lamb diffusing freely. It is natural to ask about the survival probability of the lamb not being captured by any of the lions. Here, due to the one-dimensional geometry of the system, the lamb will be caught by the rightmost lion, whose trajectory is
\begin{equation}
    x_{\rm lead}(t) = \max_{i} x_i(t) \;,
\end{equation}
where the subscript `lead' stands here for the `leader'. Of course, the lions might change their order, and one of them may be the leader of the pride for some time until it is overtaken by another. Note that in this case, since each $x_i(t)$ undergoes Brownian diffusion with diffusion constant $D$, the distribution of $x_i(t)$ at a fixed time $t$ is a Gaussian random variable as in \cref{eq:p_Gauss} with $\sigma \equiv \sigma_{\rm BM} = \sqrt{2 D\,t}$. Computing the survival probability of the lamb for any finite $N$ is a notoriously difficult problem~\cite{bray2013persistence}.
However, analytical progress can be made in the limit of large $N \gg 1$ by exploiting 
the results just mentioned for the EVS of Gaussian random variables.  
Specifically, at fixed $t$, the distribution of $x_{\rm lead}(t)$ converges, as $N \to \infty$, to a Dirac delta function centered at $a_N \approx \sqrt{2Dt \, \ln N}$, and the leader's  trajectory $x_{\rm lead}(t)$ becomes deterministic, namely, 
\begin{equation}
    x_{\rm lead}(t) \approx \sqrt{2Dt \ln N} \;.
    \label{eq:last_lion}
\end{equation}
Therefore, in this limit $N \gg 1$, the survival probability of the lamb not being captured by the lions coincides with the one obtained for a Brownian particle in the presence of an absorbing boundary moving deterministically according to \cref{eq:last_lion}. Interestingly, this simpler problem can eventually be solved, and the survival probability of the lamb can be obtained explicitly in the large $N$ limit. We refer the interested reader to Refs. \cite{kabluchko2011extremes,le2024dynamics,majumdar2024decorrelation} for refined studies of the extremal process $x_{\rm lead}(t)$ beyond this simple deterministic description in Eq. (\ref{eq:last_lion}). 

\vspace*{0.5cm}
\noindent{\bf The Fr\'echet universality class II.}
Next, let us consider the case of a parent distribution again with an unbounded support, but now characterized by an integrable power-law tail
\begin{equation}
    p(x) \approx \frac{A}{x^{\alpha + 1}}  \quad \mbox{with} \quad   A,\alpha>0 \quad \mbox{as} \quad x \to \infty \;.
\end{equation}
This is the so-called Fr\'echet universality class. In this case, the cumulative distribution function and the corresponding probability distribution function are of the form
\begin{equation}
    G_{\mathrm{II}}(y) = \begin{cases}
        e^{-y^{-\alpha}}\;, \quad \, \,  \mbox{for} \quad y>0 \;, \\ 0\;,  \qquad \quad  \mbox{for}\quad y<0\;, 
    \end{cases}
    \quad \quad g_{\mathrm{II}}(y) = \begin{cases}
\frac{\alpha}{y^{\alpha+1}}e^{-y^{-\alpha}}\;, \quad \, \, \,  \mbox{for} \quad y>0 \;,   \\
0,  \qquad \qquad  \quad  \,  \mbox{for}\quad y<0 \;,
    \end{cases} 
    \label{eq:frechet_dist}
\end{equation}
while the centering and scaling coefficients are given by
\begin{align}
\begin{cases}    
    a_N = 0\;, \\ 
    b_N = \mu_N\;,
    \end{cases}
    \label{eq:a_n_frechet}
\end{align}
with $\mu_N$ defined in \cref{eq:typ_val_def}. In particular, for large $N$, it behaves as $\mu_N \approx (N\,A/\alpha)^{1/\alpha}$. 
We note that the resulting PDF of the maximal value follows a power-law with the same exponent as the parent distribution, i.e., $g_{\rm II}(y) \propto 1/y^{\alpha+1}$, as it can be read off from Eq.~(\ref{eq:frechet_dist}).

\vspace*{0.5cm}
\noindent{\bf The Weibull universality class III.}
Finally, let us consider the third and last case of the so-called Weibull universality class. It concerns time series of  IID variables for which the parent distribution has a bounded support that behaves as
\begin{align}
    &p(x)  = 0\;, \qquad  \qquad \quad \, \, \, \, \mbox{for}\quad x>X^*\;, \\ &p(x) \sim (X^* -x)^{\alpha-1}\;,   \quad \mbox{as} \quad x \to X^{*-}\;,
    \label{eq:Wiebull_parent}
\end{align}
where $\alpha>0$. The resulting expressions for the rescaled CDF and PDF are
\begin{equation}
    G_{\mathrm{III}}(y) = \begin{cases}
        1\;, \qquad \quad  \mbox{for} \quad y>0\;, \\ e^{-|y|^{\alpha}}\;,   \quad \, \, \mbox{for}\quad y<0\;, 
    \end{cases}
    \quad g_{\mathrm{III}}(y) = \begin{cases}
0\;,  \qquad \qquad  \qquad  \mbox{for} \quad y>0\;,   \\
\alpha |y|^{\alpha-1}e^{-|y|^{\alpha}}\;, \quad \, \, \mbox{for}\quad y<0\;,
    \end{cases} 
    \label{eq:weibull_dist}
\end{equation}
while in this case, the centering and scaling coefficients are given by
\begin{align}
\begin{cases}    
    a_N = X^*\;, \\ 
    b_N = X^* - \mu_N\;,
    \label{eq:a_n_weibull}
    \end{cases}
\end{align}
where $\mu_N$ is defined in \cref{eq:typ_val_def}. 

\vspace*{0.5cm}

\begin{figure}
    \centering
    \begin{tabular}{cc}
        {\includegraphics[width=0.45\linewidth]{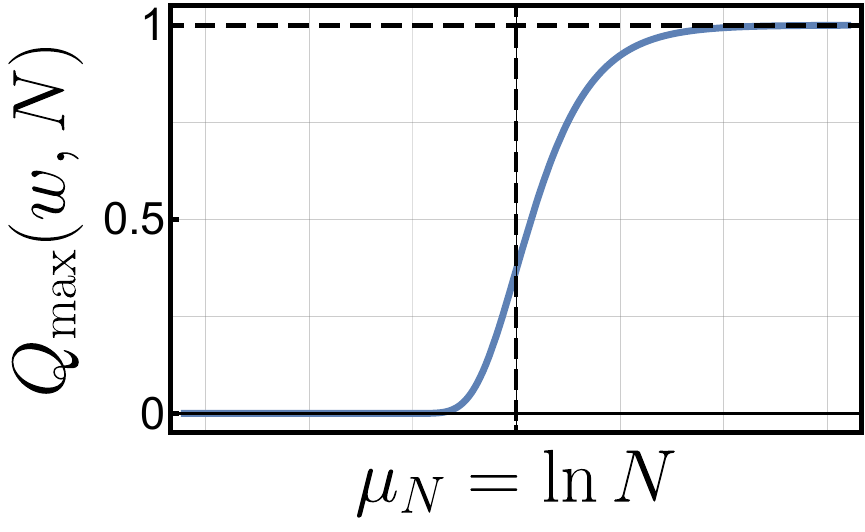}}&
        {\includegraphics[width=0.45\linewidth]{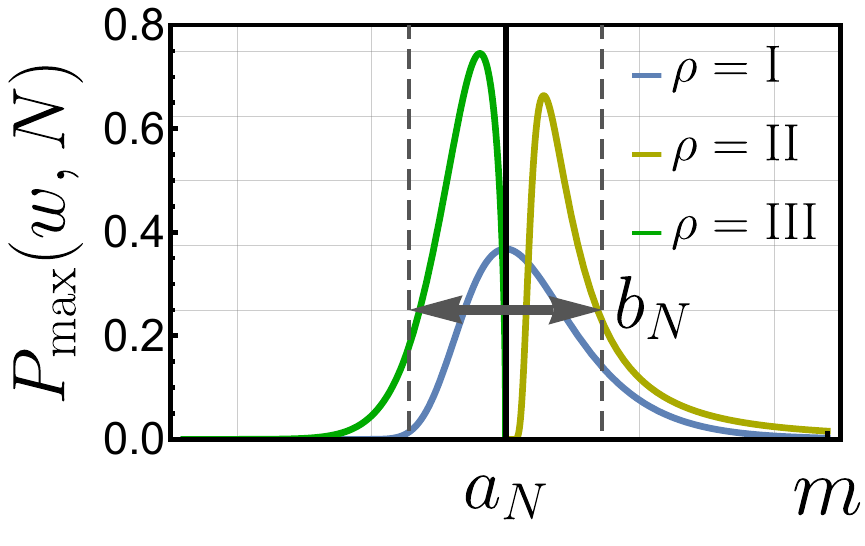}}
        \\[-2mm]
         (a) & (b) 
    \end{tabular}
\caption{In  panel (a), we plot the CDF of the maximum of a time series obtained {for $N=2000$} IID variables distributed exponentially \eqref{eq:p_exp} which  fall into the Gumbel universality class as $N \to \infty$.  Notice its sigmoidal shape, centered at $w = \ln N$ corresponding to the typical value of $X_{\max}$, obtained from \cref{eq:typ_val_def}. In panel (b), we plot the PDFs of the maximum of IID variables falling into the Gumbel, Fr\'echet and Weibull universality classes for $\rho$ = I, II, III respectively. The corresponding expressions for the limiting PDFs are given in Eqs.~\eqref{eq:gumbel_dist}, \eqref{eq:frechet_dist} and \eqref{eq:weibull_dist}. 
Note that the position of the bell-shaped curves is set by $a_N$, whereas their width is determined by $b_N$ -- see Eq.~(\ref{eq:scaled_var}).}
\label{fig:EVS_class_example}
\end{figure}

In \cref{fig:EVS_class_example}(b), we show a plot of the resulting PDFs for all three universality classes: Gumbel (I), Fr\'echet (II), and Weibull (III). For the last two cases (II and III), we chose $\alpha = 3/2$.
\subsection{Weakly correlated random variables}
We conclude the discussion of the EVS of IID variables by generalizing the results to the case of weakly correlated variables. Let us suppose that the correlations between the entries of a time series $\{x_1, x_2, \ldots, x_N \}$  are of the form 
\begin{equation}
    C_{i,j} = \langle x_i x_j \rangle - \langle x_i\rangle \langle x_j \rangle \sim e^{-|i-j|/\xi}\;,
    \label{eq:corr_length_def}
\end{equation}
where  $\xi$ is the  resulting correlation length. In the regime of $N \gg \xi$, one can \textit{coarse grain} the system by introducing blocks of length $\xi$.  We can then consider the local maxima of each of these blocks \cite{Majumdar_2020_EVS_review}
\begin{equation}
    \tilde{x}_i = \max\{x_{i\xi+1}, x_{i\xi+2}, \ldots, x_{(i+1)\xi }\}\;, \qquad i = 0,1, 2, \ldots N' = N/\xi - 1
\end{equation}
and notice that the maximum of these local maxima is the global maximum of the original time series
\begin{equation}
    X_{\mathrm{max}} = \max\{x_{1}, x_{2}, \ldots, x_{N}\} = \max\{\tilde x_{0}, \tilde x_{2}, \ldots, \tilde x_{N'}\}\;.
\end{equation}
However, the entries of the coarse-grained time series $\{\tilde x_0, \tilde x_2, \ldots, \tilde x_{N'} \}$ are essentially uncorrelated. Then, after assessing the parent distribution of the coarse-grained entries $\tilde x_i$'s, one can simply use the results known for IID variables described in Section~\ref{subsec:universality_classes}. 
Note also that in the case of Gaussian random variables $x_i$'s with stationary correlations, i.e., $C_{i,j} = c(|i-j|)$ in Eq. (\ref{eq:corr_length_def}), it has been proved rigorously \cite{berman1964limit} that if $c(x) \ll 1/\ln(x)$ as $x \to \infty$, then the limiting distribution of $X_{\max}$, properly centered and scaled, remains of the Gumbel form as $N \to \infty$. This indeed shows that the results for the EVS of IID random variables are actually quite robust and have a much wider range of validity than the strict IID case.

As a nice and simple application of the EVS of weakly correlated random variables, we discuss here the maximum of the 
Ornstein-Uhlenbeck (OU) process. This process describes the dynamics of a Brownian particle in a one-dimensional harmonic potential, evolving via Langevin dynamics
\begin{equation}
    \frac{\mathrm{d}x}{\mathrm{d}t} = -\mu x +\eta(t),
\end{equation}
where $\mu$ is the strength of a harmonic potential with its minimum at the origin, $\eta$ is a Gaussian white noise with a vanishing mean and variance $\langle \eta(t)\eta(t') \rangle = D \delta(t-t')$, where $D$ is the diffusivity. We assume that the process is initialized at the origin $x(0)=0$. Then, $x(t)$ is a Gaussian process, which is fully characterized by its two-point function
\begin{equation}
    \langle x(t_1)x(t_2) \rangle = \frac{D}{\mu} \left(e^{-\mu|t_1 - t_2|}-e^{-\mu(t_1 + t_2)} \right).
\end{equation}
In the regime of $t_1, t_2 \gg 1/\mu$, keeping $|t_1 - t_2|$ fixed (corresponding to the stationary dynamics of the OU process), 
the correlation function decays exponentially $\langle x(t_1)x(t_2) \rangle \approx D/\mu \, e^{-\mu |t_1-t_2|}$, with $1/\mu$ playing the role of the correlation length $\xi$ in \cref{eq:corr_length_def}. This is thus a continuous-time version of the case discussed above -- see Eq. (\ref{eq:corr_length_def}) -- which shows that the variables $\{x(\tau)\}$ with $\tau \in [0,t]$ form a weakly correlated (continuous) time series. Therefore, we expect that the limiting CDF of the maximal value $X_{\max}(t) = \max_{0 \leq \tau \leq t} \{x(\tau)\}$, properly centered and scaled, converges to the Gumbel form \eqref{eq:gumbel_dist} as $t \to \infty$.  
In fact, the CDF $Q_{\max}(w,t)$ of $X_{\max}(t)$ can be computed explicitly~\cite{Majumdar_2020_EVS_review}, and in the limit $t \to \infty$, it can be shown that indeed 
\begin{equation}
    Q_{\max}(w,t) \sim G_{\rm I}\left[\sqrt{4\mu \ln t}\left(w - \sqrt{\frac{\ln t}{\mu}}\right) \right],
\end{equation}
where $G_{\rm I}$ is the Gumbel distribution \eqref{eq:gumbel_dist}. 
We refer the reader to Ref. \cite{Majumdar_2020_EVS_review} for more details on the EVS of the OU process, as well as to the more recent works \cite{mori2021distribution,mori2022time} for the study of other extreme observables of the stationary OU process, such as the time at which the maximum is reached.

\section{EVS of time series -- Random Walks and Brownian Motion}
\label{sec_RW}
An important class of stochastic processes are random walks (RWs), which generically exhibit strong correlations. Before discussing their survival probabilities in \cref{sec:RWs_survival}
— closely connected to the EVS, as discussed in the introduction — we provide
a short introduction to the topic of RWs, 
 describing their properties relevant to
 the study of the statistics of extremes.
\subsection{Brief Introduction to Random Walks}
Let us define a random walk as a discrete-time stochastic process evolving via the simple Markov rule
\begin{equation}
    x_n = x_{n -1} + \eta_n \;, \qquad x_0 =0 \;,
    \label{eq:RW_def}
\end{equation}
where $\eta_n$'s are IID jumps (steps) random variables. For simplicity, we restrict ourselves  
to the case where their PDF $p(\eta)$ is continuous and symmetric, i.e., $p(\eta) = p(-\eta)$. We notice that $x_n$ is simply the sum of $n$ jumps, i.e., $x_n = \sum_{k=1}^n \eta_k$. Hence,
\begin{align}
        &\left \langle x_n \right \rangle = 0 \;, \\ 
        &\left \langle x_n x_n' \right \rangle = \sum_{k=1}^n\sum_{k'=1}^{n'} \langle \eta_k \eta_{k'} \rangle = \sigma^2 \min(n, n') \;,
\end{align}
where we used $\langle \eta_n\rangle = 0$, which follows from the symmetry of $p(\eta)$, and $\langle \eta_n \eta_{n'}\rangle = \sigma^2\delta_{n,n'}$,  assuming that the  PDF of the jumps has a finite second moment. 
We note that the  two-point function of $x_n$ differs markedly from that of the weakly correlated case in Eq.~(\ref{eq:corr_length_def}), which shows that the variables $x_n$ and $x_{n'}$ are strongly correlated even for $n \gg n'$.

Depending on the tail of the jump distribution $p(\eta)$, we distinguish two main classes of RWs: (i) if $p(\eta)$ has a finite second moment $\sigma^2 = \langle \eta^2 \rangle$, the resulting process $x_n$ is a standard random walk that exhibits a normal diffusive behavior $x_n \sim \sqrt{n}$ at large $n$; (ii) if for large $\eta$, $p(\eta)$ is characterized by a power-law decay of the form $p(\eta) \sim 1/\eta^{\mu + 1}$ with the L\'evy index $\mu\in]0, 2[$, its second moment diverges. Then, the resulting random walk is called a L\'evy flight~\cite{BG90,MK00}, which exhibits a super-diffusive behavior $x_n \sim n^{1/\mu}$ for $n \gg 1$. We compare side-by-side the main properties of standard random walks and L\'evy flights in  \cref{tab:RWs}. 
\begin{table}
\begin{center}
\begin{tabular}{ || c || c  || } 
  \hline
   \multirow{2}{*}{Standard Random Walk}  
  & \multirow{2}{*}{L\'evy flight}   \\
  & \\ 
  \hline
   \multirow{2}{*}{$\langle \eta_k^2 \rangle = \sigma^2 < +\infty$} &  \multirow{2}{*}{$p(\eta) \sim 1/\eta^{\mu +1}$ for large $\eta$  with $2>\mu>0$} \\ & \\ 
   \hline
   \multirow{2}{*}{$x_n \sim \sigma \sqrt{n}$  for large $n$} & \multirow{2}{*}{$x_n \sim n^{1/
   \mu}$  for large $n$}\\  & \\ 
  \hline
  \multirow{3}{*}{$x_n/\sigma \sqrt{n}$  converges to Brownian Motion} & \multirow{2}{*}{$x_n/ n^{1/\mu} \xrightarrow[n \to \infty]{} F_{\mu}$ (L\'evy stable dist.)} \\  & \\  &  \multirow{2}{*}{$\int_{\mathbb{R}} \mathrm{d}x \, e^{ikx}F_{\mu}(x) = e^{-|k|^{\mu}}$ }\\ \multirow{3}{*}{$x_n/\sigma \sqrt{n} \xrightarrow[n \to \infty]{} \mathcal{N}(0, 1) \quad$ (normal dist.)} & \\  & \multirow{2}{*}{$F_{\mu}(x) \sim 1/x^{1 + \mu}$ for large $x$ } \\  &  \\   
  \hline
   \multirow{2}{*}{An example of a trajectory} & \multirow{2}{*}{An example of a trajectory} \\ 
    &   \\ 
  \begin{minipage}{.45\textwidth}
      \includegraphics[width=\linewidth]{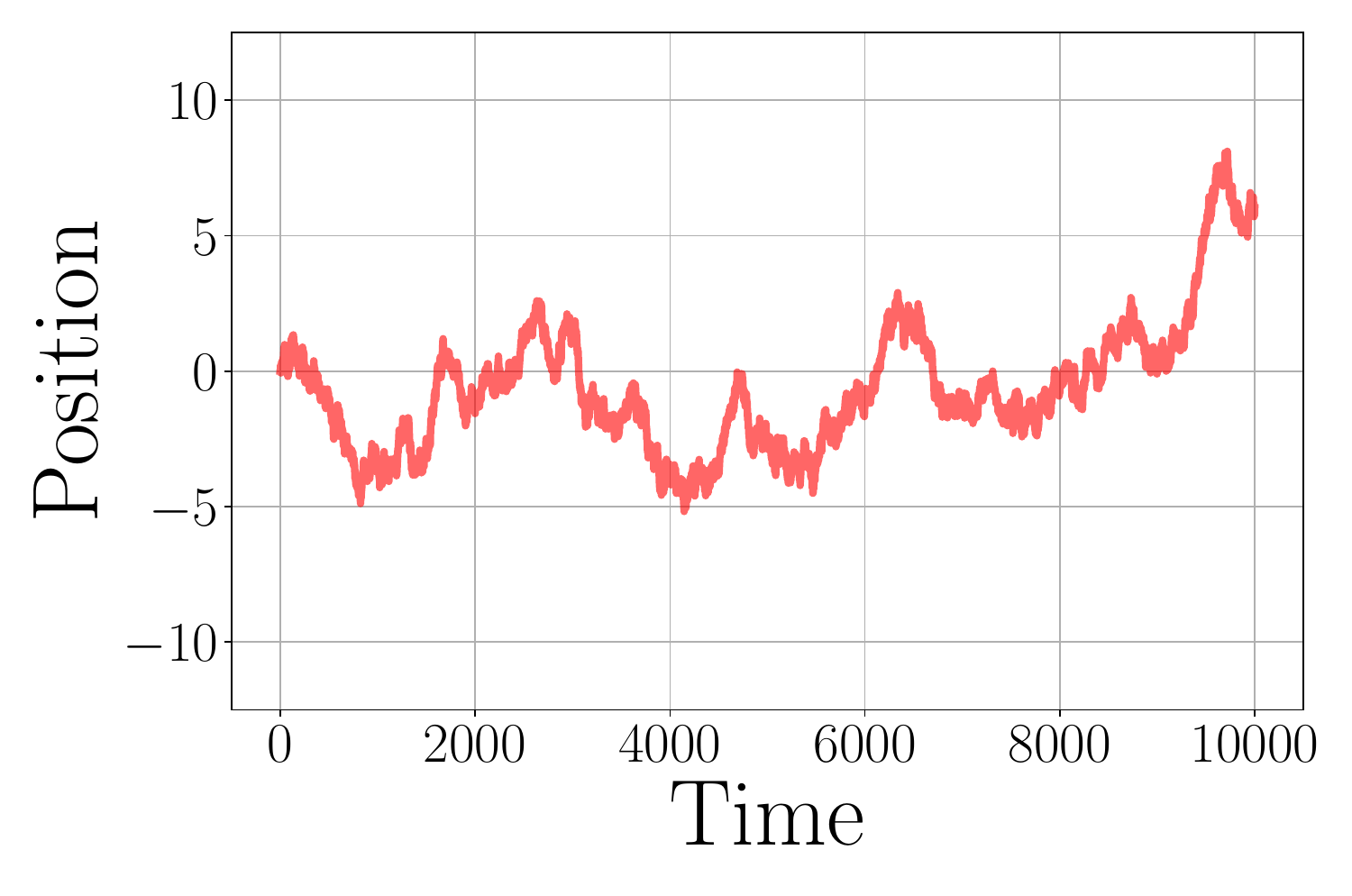}
    \end{minipage} & \begin{minipage}{.45\textwidth}
      \includegraphics[width=\linewidth]{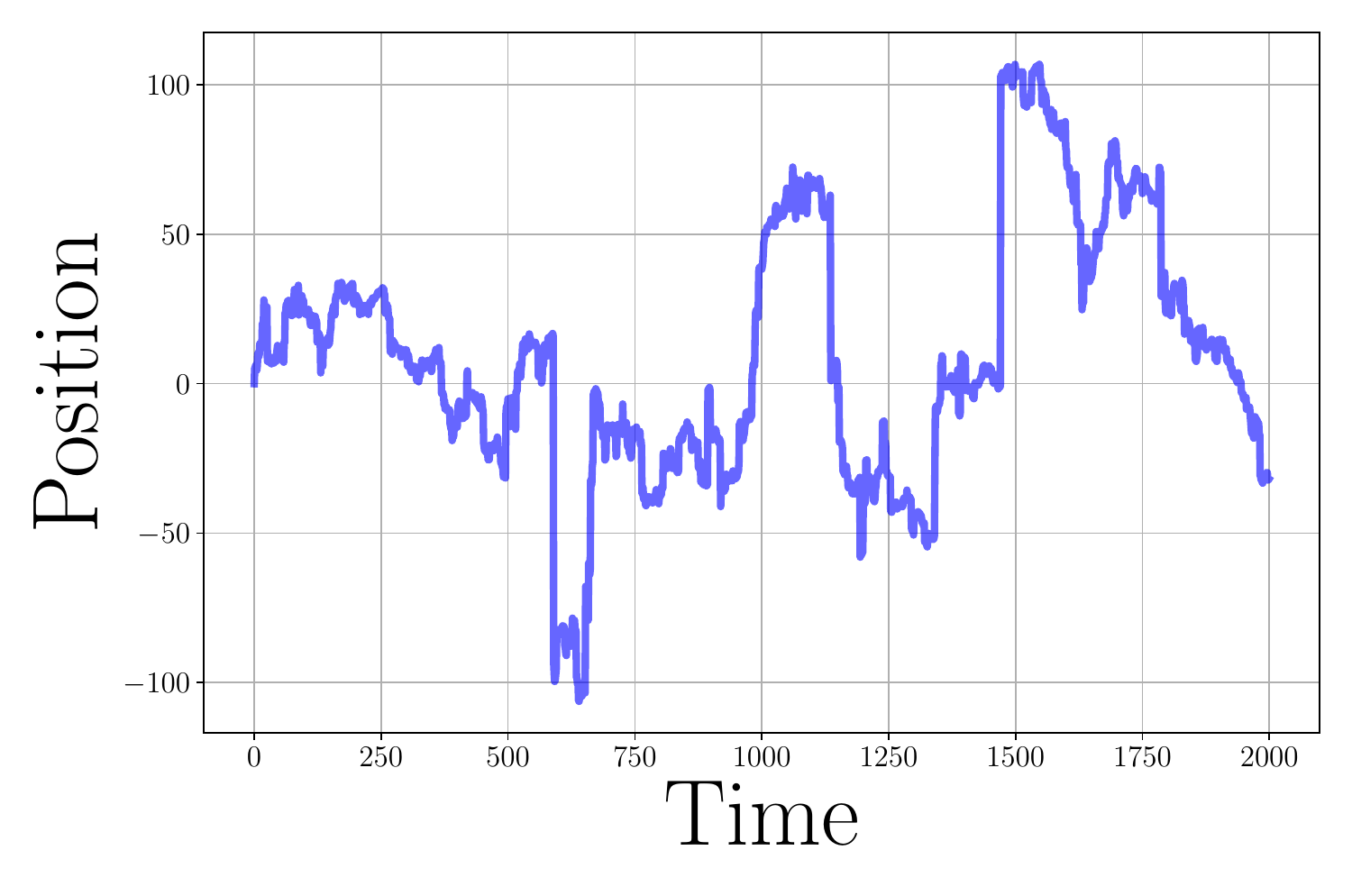}
    \end{minipage} \\ 
    \hline
\end{tabular}
\end{center}
 \caption{Two major classes of random walks, standard random walks vs L\'evy walks, and their salient features.}
 \label{tab:RWs}
\end{table}

To describe the statistical properties of a random walk, it is useful to introduce the Green's function of the process $G(x, x_0|n)$, which is the PDF
that the walker's position is within the interval $[x, x + \mathrm{d}x]$ after the $n$-th step, assuming that it started its motion at $x_0$. In particular, by definition, it satisfies the initial condition $G(x, x_0|0) = \delta(x -x_0)$. Due to Markovianity of the process (\ref{eq:RW_def}), i.e., the fact that the state of the process after $n$ steps depends solely on its state after $n-1$ steps and carries no memory of the previous dynamics, we can conveniently describe the evolution of the Green's function in terms of the \textit{forward Fokker-Planck equation}, namely
\begin{equation}
    G(x, x_0|n) = \int_{\mathbb{R}}\mathrm{d}x' \, G(x', x_0|n-1) p(x - x') \;. 
    \label{eq:for_FP}
\end{equation}
This equation simply accounts for the event of the particle {jumping} from the position $x'$ at step $n-1$ to the current position $x$ at step $n$ by an amount $(x-x')$, which is drawn randomly from the distribution $p(x-x')$. This is called a `forward' equation because one considers the current position $x$ as a random variable, while the initial position $x_0$ is kept fixed.

Though the forward equation (\ref{eq:for_FP}) is very natural, one can also write down a {\it backward} equation where one considers the starting position $x_0$ as the variable, while the final position $x$ is kept fixed. The {\textit{backward Fokker-Planck equation}}
then reads
\begin{equation}
    G(x, x_0|n) = \int_{\mathbb{R}}\mathrm{d}x_0' \, G(x, x_0'|n-1)p(x'_0 - x_0) \;.
    \label{eq:back_FP}
\end{equation}
This equation is obtained by considering the displacement of the particle at the first step from $x_0$ to $x'_0$ and then taking into account the subsequent evolution up to $(n-1)$ steps, but now starting from $x'_0$. In the context of first-passage problems, the backward equation (\ref{eq:back_FP}) is usually computationally advantageous \cite{bray2013persistence,Majumdar10}. Of course, both approaches allow us to solve for the Green's function, leading to the same results. This can be conveniently done via the use of the Fourier transform, namely  
\begin{equation}
    G(x, x_0|n) = \int_{\mathbb{R}}\frac{\mathrm{d} k}{2 \pi}e^{-ik (x - x_0)} \left[\hat p(k) \right]^n,
\end{equation}
where we defined the Fourier transform $\hat p(k)$ of the jump distribution via
\begin{equation}
    \hat p(k) = \int_{\mathbb{R}}\mathrm{d} x \,e^{ikx}\,p(x)\;.
    \label{eq:jump_dist_FT}
\end{equation}
We remark that in the case where $p(x)$ has a finite second moment $\sigma^2$, one can obtain the continuum limit of Eq. (\ref{eq:for_FP}) -- or equivalently of Eq. (\ref{eq:back_FP}) -- by introducing $\Delta t$ and $D>0$ so that
\begin{equation} \label{eq:continnum}
    t = n \Delta t \;, \qquad \sigma^2 = 2 D \Delta t \;.
\end{equation}
Then, in the limit of $\Delta t \to 0$, the forward Fokker-Planck equation in Eq.~\eqref{eq:for_FP} converges to the well known 
diffusion equation 
\begin{equation}
    \frac{\partial G(x, x_0| t)}{\partial t} = D\frac{\partial^2 G(x, x_0| t)}{\partial x^2} \;,
    \label{eq:diff}
\end{equation}
while the backward Fokker-Planck equation in Eq.~\eqref{eq:back_FP} converges to the backward diffusion equation 
\begin{equation}
    \frac{\partial G(x, x_0| t)}{\partial t} = D\frac{\partial^2 G(x, x_0| t)}{\partial x_0^2} \;,
    \label{eq:diff_back}
\end{equation}
In both cases, the initial condition reads $G(x,x_0 \vert t= 0) = \delta(x-x_0)$.

It is interesting to note that the discrete Fokker--Planck equations \eqref{eq:for_FP} and \eqref{eq:back_FP} take the mathematical form of 
integral equations. In the continuum limit, by contrast, the dynamics of the Green's function is governed by partial differential equations, see Eqs.~\eqref{eq:diff} and \eqref{eq:diff_back}. For L\'evy flights with 
$0<\mu<2$, for which $\sigma^2$ diverges, these equations are replaced 
by a L\'evy fractional diffusion equation (see, e.g.,~\cite{MK00} for 
a review and discussion).

\subsection{Survival probability and the first passage times}  
\label{sec:RWs_survival}

\begin{figure}
    \centering
    \includegraphics[width=1\linewidth]{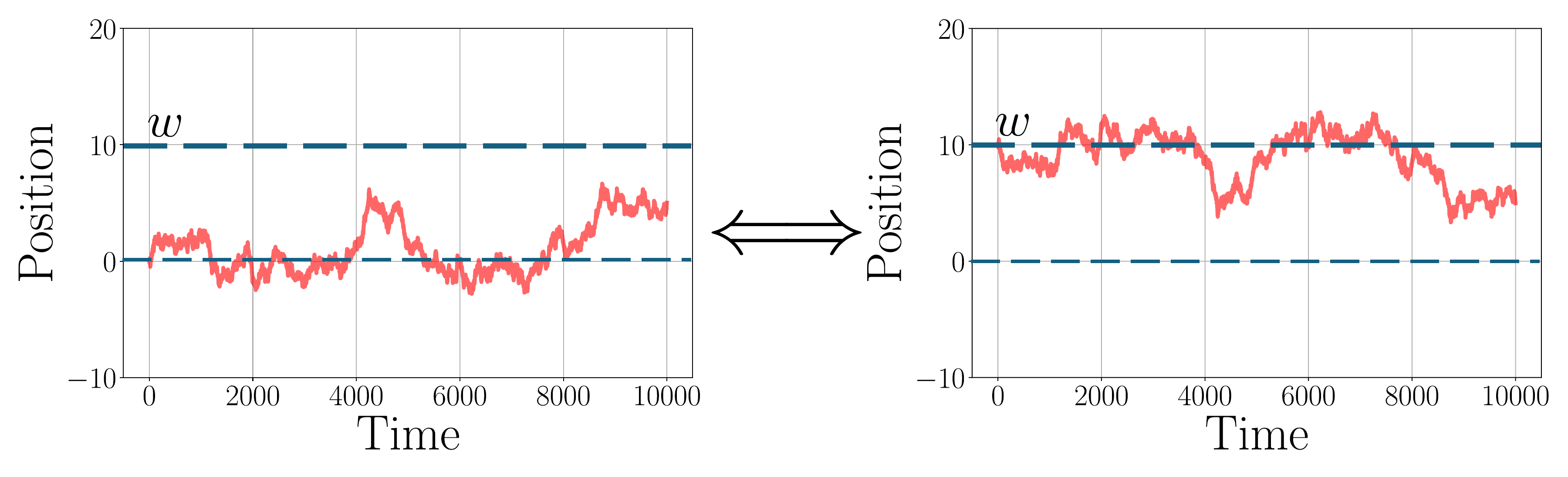}
    \caption{Due to the fact that the distribution of jumps $p(\eta)$ is even under $\eta \to -\eta$, and does not depend on time, probabilities associated with a trajectory and a trajectory obtained by its reflection in space or time are the same.  Therefore,   the probability that the value of a random walk beginning at the origin remains below $w>0$ is equivalent to the probability of it remaining above $-w$. Moreover, because the system is invariant under spatial translations, this probability is equal to the one associated with a random walker \textit{beginning its motion at $w$} and staying above $0$. 
    }
    \label{fig:stay_positive}
\end{figure}

Let us now turn our focus to the survival probability of a random walker staying below $w$ after $n$ steps, which, as mentioned in the introduction, coincides with the CDF of the maximum $Q_{\max}(w,n)$ (see Fig.~\ref{fig:phys_interpret}). Indeed, because of the translational invariance of the system and the symmetry of the jump distribution under $x \mapsto -x$, evaluating $Q_{\max}(w,n)$ of a process initialized at $x_0 =0$ is equivalent to solving for the survival probability of the same process \textit{starting} at $w>0$ and staying above $0$ after $n$ steps. We denote the latter survival probability by $Q(w, n)$. The equivalence of the two problems is illustrated in \cref{fig:stay_positive}. 

In order to solve for $Q(w, n)$, we first introduce the restricted Green's function
\begin{equation}
    G_+(x, x_0|n) = {\rm Prob.}(x_1>0, x_2>0, \ldots, x_{n-1}>0,  x_n \in [x, x+ \mathrm{d}x]\big |x_0>0) \;,
\end{equation}
which denotes the probability that the walker's position is within the interval $[x, x + \mathrm{d}x]$ -- with $x>0$ -- after $n$ steps \textit{and} that its trajectory remained on the positive half-line. Its dynamics is given by
\begin{align}
     G_+(x, x_0|n) = \int_0^{\infty}\mathrm{d}x' \, G_+(x', x_0|n-1) p(x - x')\;, \qquad  \quad \mbox{(forward)}, \\  G_+(x, x_0|n) = \int_{0}^{\infty}\mathrm{d}x_0' \, G_+(x, x_0'|n-1) p(x_0 - x_0')\;, \qquad  \mbox{(backward)}.
     \label{eq:rest_greens}
\end{align}
Note that, in contrast with Eqs.~\eqref{eq:for_FP} and \eqref{eq:back_FP}, the integrals on the right-hand side of the equations above are over $\mathbb{R}_+$ instead of the whole $\mathbb{R}$. This is because the random walker is constrained to stay positive. 
Both equations above are valid for $x \geq 0$ and  $x_0 \geq 0$ and both satisfy the initial condition $G_+(x, x_0 |0) = \delta(x - x_0)$. For the study of the survival probability, it turns out to be more convenient to deal with the backward dynamics in \cref{eq:rest_greens}. Indeed, by integrating over $x \in [0, \infty[$ on both sides of Eq. (\ref{eq:rest_greens}), one immediately obtains a closed integral equation for the survival probability, which reads
\begin{equation}
    Q(x_0, n) = \int_{0}^{\infty}\mathrm{d}x_0' \, Q(x_0', n-1) p(x_0 - x_0') \;,
    \label{eq:RW_surviva_int_eq}
\end{equation}
together with the initial condition
\begin{equation}
    Q(x_0, 0) = 1, \qquad \mbox{for} \quad x_0 > 0 \;.
\end{equation}
The integral equation for the survival probability \eqref{eq:RW_surviva_int_eq} is not easy to solve generically. However, assuming that the distribution of the jumps $p(\eta)$ is continuous, a closed form for the Laplace transform of its generating function can be obtained. It is given by the so-called Pollaczek-Spitzer formula \cite{Pollaczek_1952, Spitzer_1956, Spitzer_1957} 
\begin{equation}
    \int_0^{\infty}\mathrm{d}x_0 e^{-\lambda x_0} \sum_{n=0}^{\infty}Q(x_0, n) s^n = \frac{1}{\lambda \sqrt{1 -s}} \exp \left[-\frac{\lambda}{\pi} \int_0^{\infty}\mathrm{d}k \frac{\ln(1 -s\hat p(k))}{\lambda^2 + k^2}\right],
    \label{eq:spitzer}
\end{equation}
where $\hat p (k)$ is the Fourier transform of the jump distribution given by \cref{eq:jump_dist_FT}. A rather straightforward yet insightful consequence of the Pollaczek-Spitzer formula is the 
Sparre Andersen theorem \cite{Sparre_Andersen_1953, Sparre_Andersen_1954} 
for $Q(x_0=0,n)$, which can be obtained by taking the limit $\lambda \to \infty$ on both sides of Eq.~(\ref{eq:spitzer}) -- see Ref. \cite{Majumdar_Schehr_book} for details. It reads
\begin{equation}
    \sum_{n=0}^\infty Q(0, n)s^n = \frac{1}{\sqrt{1-s}} \;.
    \label{eq:spitzer_0}
\end{equation}
Quite surprisingly, this result does not depend on the PDF of the jumps $p(\eta)$ and is, in that sense, universal. Expanding the right-hand side of \cref{eq:spitzer_0} in powers of $s$ and comparing the coefficients term by term gives the expression for the survival probability for a walker starting at $x_0  = 0$ in a fully explicit form
\begin{equation}\label{eq:SA_explicit}
    Q(0,n) = \binom{2n}{n}2^{-2n} \;,
\end{equation}
which is \textit{universal}, i.e., independent of the jump distribution $p(\eta)$, \textit{for all $n$}, and not just for large $n$, where it decays as $Q(0,n) \approx 1/\sqrt{\pi \,n}$.

\begin{figure}
    \centering
    \begin{tabular}{cc}
        {\includegraphics[width=0.55\linewidth]{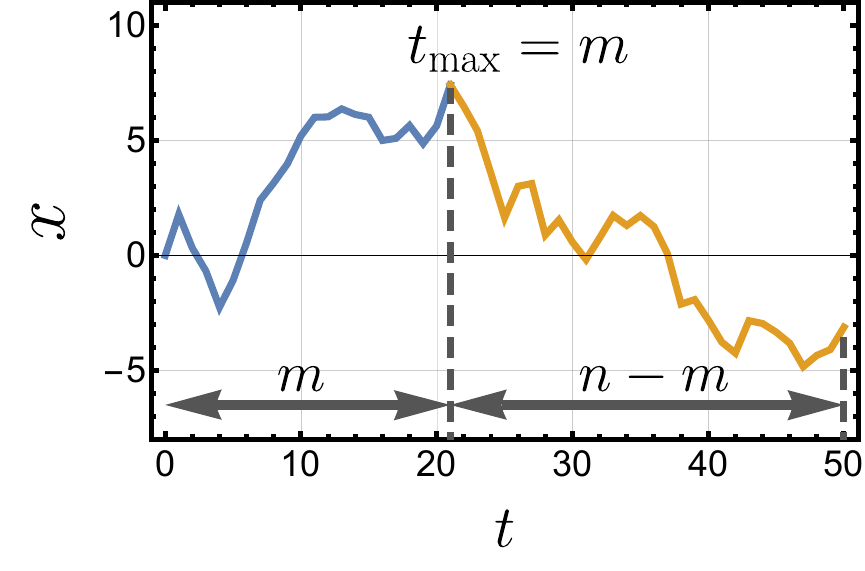}}&
    \end{tabular}
\caption{A trajectory of an $n$-step random walker is split into two parts: one from the beginning of its motion to the step $t_{\rm max} = m$ at which it attains its maximal value (the blue curve),   and one from the $m$-th step to its final step  (the yellow curve).
Due to Markovianity of the process, these two segments are statistically independent.}
\label{fig:Sparre_Andersen}
\end{figure}

The fact that the result \eqref{eq:spitzer_0} is independent of the specific form of $p(\eta)$ suggests that it may admit a simple derivation. 
This is indeed the case \cite{dembo2013persistence,mounaix2020statistics,Majumdar_Schehr_book}, and we briefly outline the argument below. 
To this end, we consider the probability that the maximum of an $n$-step random walk, which may take negative values, occurs at the $m$-th step. 
We denote this probability by
\begin{equation}
    P(m|n) = \mathrm{Prob.}(t_{\max}=m \mid n).
    \label{eq:prob_tmax}
\end{equation}
We can now separate the trajectory of the random walker into two segments: the first runs from the beginning of the motion to the $m$-th time step, at which the process reaches its maximal value, and the second extends from this $m$-th step to the final step $n$, as sketched in \cref{fig:Sparre_Andersen}. 
Owing to the Markovian nature of the process, these two parts of the trajectory are statistically independent. 
Moreover, by the symmetry illustrated in \cref{fig:stay_positive}, the probability that a trajectory of length $m$ attains its maximum at its final position is exactly equal to the survival probability of a process starting at $x_0 = 0$ and remaining positive up to the $m$-th time step, namely $Q(0,m)$. 
This follows from the fact that each trajectory starting at $0$ whose maximum occurs at step $m$ can be uniquely mapped onto a trajectory starting at $0$ that stays positive during its entire $m$-step evolution. By the same argument, the probability that the second segment of length $n-m$ remains below the value attained at step $m$ is $Q(0, n-m)$.
Hence, by the Markov property, the probability $P(m|n)$ is given by the product of these two factors, namely
\begin{equation}
   P(m|n) = Q(0,m)Q(0,n-m) \;.
   \label{eq:Q_from_prob_tmax}
\end{equation}
We can now sum both sides of the equation above over $m\in \{0,1,2, \ldots, n\}$ and use the normalization of the probability distribution $P(m \vert n)$, which yields
\begin{equation}
    \sum_{m=0}^n P(m|n) = \sum_{m=0}^n Q(0,m)Q(0,n-m) =1 \;.
\end{equation}
Multiplying both sides by $s^n$ and summing over all $n \in \mathbb{N}$ gives 
\begin{equation}
\sum_{n=0}^{\infty}s^n \sum_{m=0}^n Q(0,m)Q(0,n-m)= \left[ \sum_{m=0}^{\infty} s^mQ(0,m)\right]^2 = \frac{1}{1-s}\;,   
\end{equation}
where we changed the summation variables from $(n,m)$ to $(m, n-m)$ in the first equality. 
The result above is equivalent to the one reported in \cref{eq:spitzer_0}.

While the Sparre Andersen result (\ref{eq:SA_explicit}) gives the survival probability starting exactly at the origin $x_0=0$, it is, in general, very difficult to compute $Q(x_0,n)$ for $x_0>0$ from the Pollaczek-Spitzer formula in Eq.~(\ref{eq:spitzer}). However, it is possible to extract the limiting form of $Q(x_0,n)$ in the scaling limit where $x_0 \to \infty$ and $n \to \infty$ with the ratio $x_0/\sqrt{n}$ fixed.
In this limit one finds  
\begin{equation} \label{eq:Q0_erf}
    Q_0(x_0, n)\approx  \erf \left({\frac{x_0}{\sigma \sqrt{2 n }}}\right), \qquad x_0 \geq 0 \;,
\end{equation}
where $\erf(x) = \frac{2}{\sqrt{\pi}}\int_0^x \mathrm{d}u \, e^{-u^2}$ is the error function. Note that this {well-known} result~(\ref{eq:Q0_erf}) can also be obtained by solving the integral equation (\ref{eq:RW_surviva_int_eq}) for $Q(x_0,n)$ in the aforementioned continuum limit (\ref{eq:continnum}), which yields $Q_0(x_0,t) = \erf{(x_0/\sqrt{4 D\,t})}$, which is indeed fully consistent with the continuum limit of Eq. (\ref{eq:continnum}).

\section{EVS in Random Matrix Theory -- Tracy-Widom laws and applications}
\label{sec_RM}
Another class of strongly correlated variables whose EVS are of importance in physical contexts consists of the eigenvalues of a random matrix (see \cite{majumdar2014top} for a review).  
Let us consider an $N\times N$ Gaussian random matrix $\mathbf{X}$ with entries $X_{i,j}$ such that the matrix is real symmetric, complex Hermitian, or quaternionic self-dual, with its entries drawn from the joint Gaussian distribution \cite{mehta2004random,forrester2010log}
\begin{equation}
    P_{\beta}[\{ X_{i,j}\}] = \frac{1}{Z_{N,\beta}}\exp\left[-\frac{\beta N }{2} \Tr \mathbf{X}^2\right],
    \label{eq:pdf_RM_entries}
\end{equation}
where $Z_{N,\beta}$ is the normalization factor, $\beta$ is the so-called Dyson index, and $\Tr$ denotes the trace of a matrix. Note that in the expression above, we scaled the entries by $\sqrt{N}$. When the Dyson index takes the value $\beta=1, 2, 4$, the random matrix belongs, respectively, to the so-called Gaussian Orthogonal Ensemble (GOE),  the Gaussian Unitary Ensemble (GUE), or the Gaussian Symplectic Ensemble (GSE). Indeed, due to the cyclic property of the trace, namely $\Tr\{ABC \}   = \Tr\{CAB \}$ for any $N \times N$ matrices $A, B$ and $C$, the distribution \eqref{eq:pdf_RM_entries} is invariant under orthogonal, unitary and symplectic transformations. Because of this symmetry, the real eigenvalues $\lambda_1, \lambda_2, \ldots, \lambda_N$ of the matrix are statistically independent of the corresponding eigenvectors. It turns out that the joint PDF of the eigenvalues takes the form~\cite{mehta2004random,forrester2010log}
\begin{equation}
    P_{\mathrm{joint}}(\lambda_1, \lambda_2, \ldots, \lambda_N) = \frac{1}{Z_{N, \beta}'} e^{-\frac{\beta N}{2} \sum_{i=1}^N \lambda_i^2}\prod_{i<j}|\lambda_i - \lambda_j|^{\beta}
    \label{eq:RM_joint_evals} \;,
\end{equation}
where $Z_{N, \beta}'$ is introduced for the sake of normalization. 
Note that the repulsive interaction term $\propto |\lambda_i - \lambda_j|^{\beta}$ gives rise to strong (anti)-correlations between eigenvalues, making it very unlikely that $\lambda_i \simeq \lambda_j$ for $i \neq j$. Although the Gaussian random matrix models defined in Eq.~(\ref{eq:RM_joint_evals}) were initially introduced and studied for the classical values $\beta=1, 2$ and $4$, it was later shown that one can actually define random (tri-diagonal) matrices whose eigenvalue distribution is given by Eq. (\ref{eq:RM_joint_evals}) for any $\beta > 0$ \cite{dumitriu2002matrix}. 

\begin{figure}
    \centering
    \begin{tabular}{cc}{\includegraphics[width=0.5\linewidth]{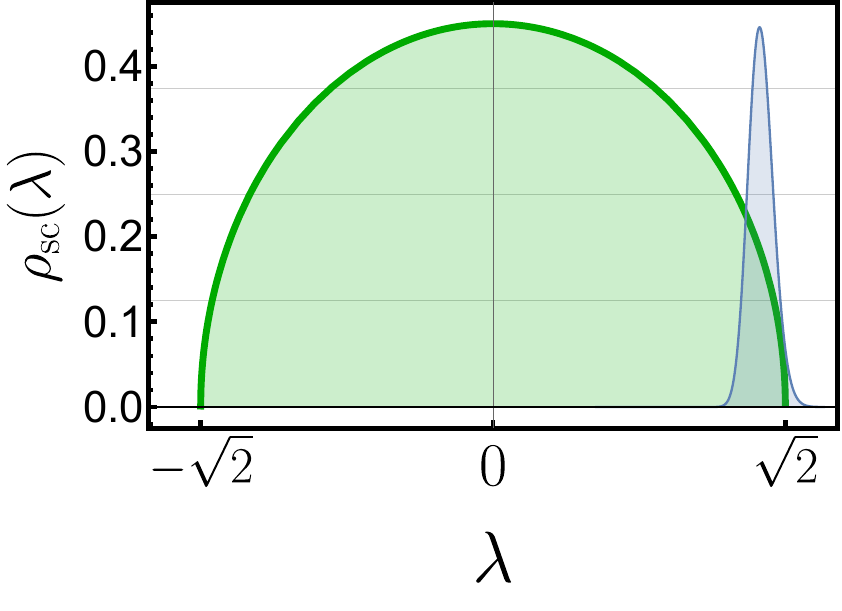}}
        &
        \raisebox{0.45 in} {{\includegraphics[width=0.45\linewidth]{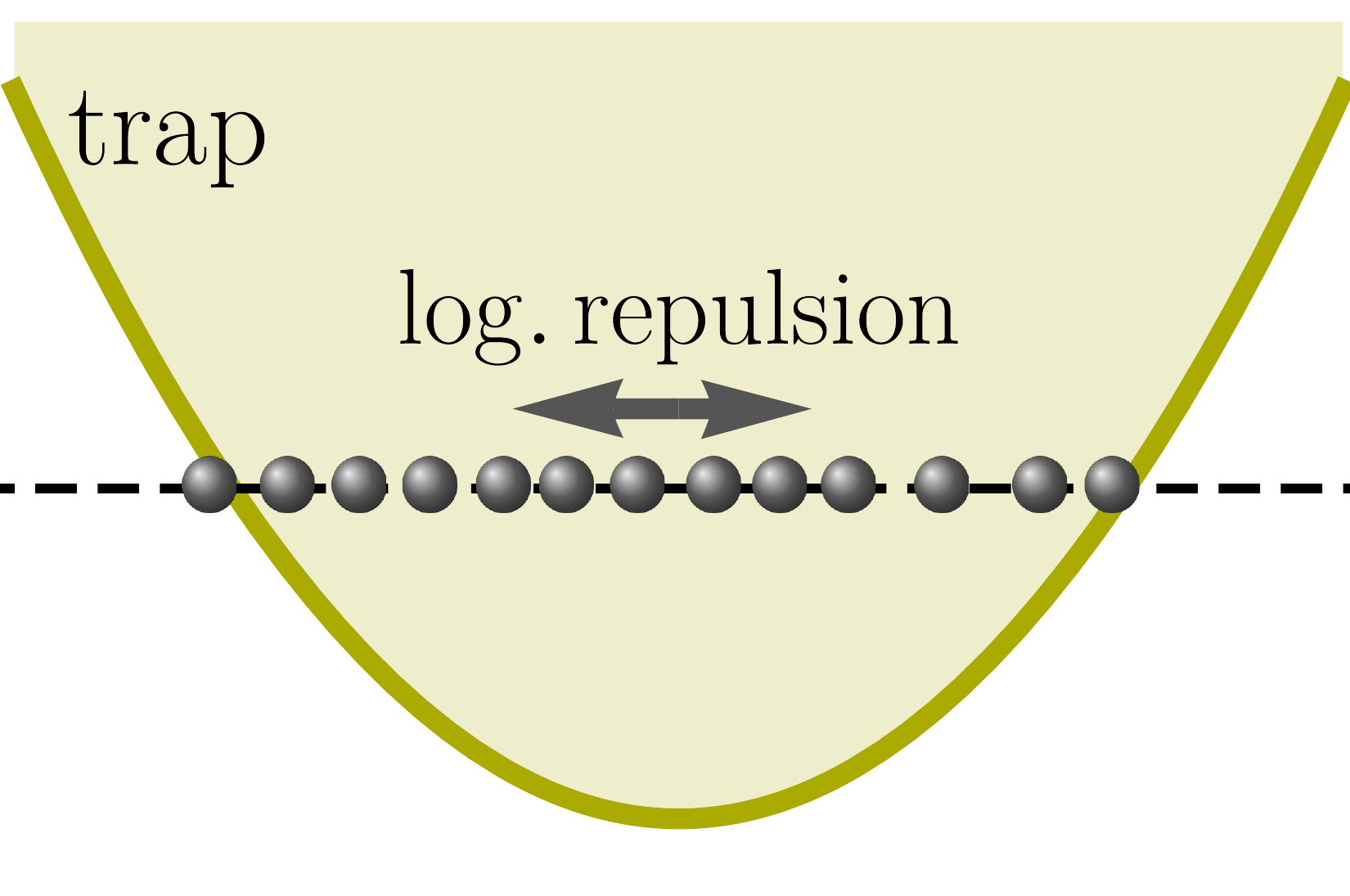}}}
        \\[-2mm]
         (a) & (b) 
    \end{tabular}
\caption{In panel (a), we plot with the green curve the Wigner semi-circle law, i.e., the empirical density of  the eigenvalues of a Gaussian random matrix in the limit of $N\to \infty$, see \cref{eq:Wigner_law}. The blue curve represents a sketch of the Tracy-Widom PDF describing the statistics of $\lambda_{\max}$ in the $N \gg 1$ regime {in the case of the GUE}. In panel~(b), we provide a cartoon of the physical system described by Dyson's log gas energy functional~\eqref{eq:Dyson_gas_energy}. Particles on a line repel each other via a logarithmic potential, causing them to spread, and are simultaneously drawn towards the origin by a harmonic trap whose strength scales linearly with $N$.}
\label{fig:Log_gas}
\end{figure}

The first observable that we want to investigate is the empirical density of the eigenvalues defined by
\begin{equation}
    \rho_N(\lambda) = \frac{1}{N} \sum_{i=1}^N \delta(\lambda - \lambda_i) \;,
\end{equation}
which is normalized to unity, i.e., $\int_{-\infty}^\infty \rho_N(\lambda) \mathrm{d} \lambda = 1$. In the large $N$ limit, it turns out that this observable is self-averaging, i.e., it becomes essentially deterministic, and converges to
\begin{equation}
    \rho_N(\lambda) \xrightarrow[N \to \infty]{} \rho_{\mathrm{sc}}(\lambda) = \frac{1}{\pi}\sqrt{2 - \lambda^2} \;,
    \label{eq:Wigner_law}
\end{equation}
which is the famous Wigner semi-circle law, a plot of which is presented in \cref{fig:Log_gas}(a). Following the seminal work of Dyson \cite{dyson1962statistical}, 
let us remark that the interaction term in \cref{eq:RM_joint_evals} can be written in the form 
\begin{equation}
    \prod_{i<j} |\lambda_i - \lambda_j|^{\beta} = e^{\beta/2 \sum_{i\neq j}\ln|\lambda_i -\lambda_j|}\;. 
\end{equation}
This allows us to express the joint probability distribution function of the eigenvalues~as
\begin{align}
    P_{\mathrm{joint}}(\lambda_1, \lambda_2, \ldots, \lambda_N) =\frac{1}{Z_{N, \beta}'}\exp \left( -\beta E[\{ \lambda_i\}]\right)   
    \end{align}
which is the Boltzmann measure of a one-dimensional gas with the energy functional of its configuration of the form
    \begin{align}
         E[\{ \lambda_i\}] = \frac{N}{2}\sum_{i=1} \lambda_i^2 - \frac{1}{2}\sum_{i\neq j}\ln|\lambda_i - \lambda_j|\;.
         \label{eq:Dyson_gas_energy}
\end{align}
This system of particles described by the energy functional above (\ref{eq:Dyson_gas_energy}) is called the Dyson's log gas~\cite{mehta2004random,forrester2010log}, which owes its name to the repulsive logarithmic interaction between the particles. 
Note that the first term in \eqref{eq:Dyson_gas_energy} describes a harmonic potential confining the particles close to the origin, whose stiffness is set by $N$. Hence, it
is of the same order as the interaction term, i.e., they both are $\mathcal{O}(N^2)$ -- since the typical scale of $\lambda_i$'s is of order ${\cal O}(1)$, as can be seen from (\ref{eq:Wigner_law}). 
Therefore, we arrived at the description of the eigenvalues of a Gaussian random matrix in terms of the thermodynamics of an interacting gas subject to a harmonic trap, represented schematically in 
\cref{fig:Log_gas}(b).

Let us return to the discussion of extreme value statistics by considering the largest eigenvalue
\begin{equation}
    \lambda_{\mathrm{max}} = \max \{\lambda_1, \lambda_2, \ldots, \lambda_N  \}\;.
\end{equation}
 From the Wigner semi-circle law in \cref{eq:Wigner_law}, one naturally expects that
\begin{equation}
    \lambda_{\mathrm{max}} \xrightarrow[N \to \infty] {} \sqrt{2} \; \;,
\end{equation}
which can be proved rigorously, see, e.g., \cite{anderson2010introduction}. In the regime of $N \gg 1$ but finite, the value of the largest eigenvalue fluctuates around $\sqrt{2} $, and one can show that the typical fluctuations scale as \cite{forrester1993spectrum,tracy1994level,tracy1996orthogonal}
\begin{equation}
\left( \sqrt{2} - \lambda_{\max} \right) \sim N^{-2/3} \;.
\end{equation}
Remarkably, although the eigenvalues are clearly correlated random variables, this scale ${\cal O}(N^{-2/3})$ can be obtained by  applying the results for IID in \cref{eq:typ_val_def} to the Wigner semi-circle law \cref{eq:Wigner_law}, see, e.g., \cite{majumdar2014top}, showing that the domain of applicability of this argument goes much beyond the IID case. In fact, the typical fluctuations of $\lambda_{\max}$ within this scale of order ${\cal O}(N^{-2/3})$ can be characterized more precisely, leading to \cite{tracy1994level,tracy1996orthogonal}
\begin{equation} \label{eq:lambda_max}
    \lambda_{\mathrm{max}} = \sqrt{2} + \frac{1}{\sqrt{2}}N^{-2/3}\chi_{\beta} + o(N^{-2/3}) \quad, \quad {\rm as} \quad N \to \infty \;,
\end{equation}
where $\chi_{\beta}$ is a Tracy-Widom random variable, which can be defined for any value of $\beta >0$ \cite{Edelman_2007_stoch_Airy, Ramirez_2011_stoch_airy} (see below for further discussions). Indeed, its CDF ${\rm Prob.}(\chi_\beta \leq x) = {\cal F}_{\beta}(x)$ is a $\beta$ Tracy-Widom law, which depends on the value of the Dyson index $\beta$. In particular, it can be shown that the corresponding PDF $\mathcal{F}_{\beta}'(x)$ has asymmetric non-Gaussian tails~\cite{BBD08,BEMN11,DV13}
\begin{equation}
\mathcal{F}_{\beta}'(x) \approx \begin{cases}
    \exp\left( -\frac{2 \beta}{3}x^{3/2}\right), \quad x \to \infty,  \\ 
    \exp \left( -\frac{\beta}{24} |x|^3 \right), \quad \, \,  x\to -\infty.
    \label{eq:TW_PDF}
\end{cases}
\end{equation}
A schematic plot of {the PDF ${\cal F}_2'(x)$} is represented by a blue curve in \cref{fig:Log_gas}(a).

Over the past few decades, it 
has been shown that the Tracy-Widom distributions appear in many different situations in mathematics and physics. For instance, they arise in combinatorics in the study of the length of the longest increasing subsequence of random permutations \cite{Baik_1999}, as well as in fundamental models of statistical physics such as interfacial growth or directed polymers in random media (see e.g. \cite{Prahofer_2000,Johansson_2000, Baik_2000,Gravner_2001,Majumdar_2004,Imamura_2004,Dotsenko_2010, Sasamoto_2010,Calabrese_2010,amir2011probability,Calabrese_2011,Imamura_2012}) -- we refer the reader to Refs. \cite{kriecherbauer2010pedestrian,corwin2016kardar} and \cite{Majumdar_Les_Houches_2007,spohn2017kardar} for recent reviews and lecture notes respectively. Below, we discuss two such emblematic occurrences of the Tracy-Widom distributions in statistical physics.

\begin{figure}[t]
    \centering
    \begin{tabular}{cc}{{\includegraphics[width=0.47\linewidth]{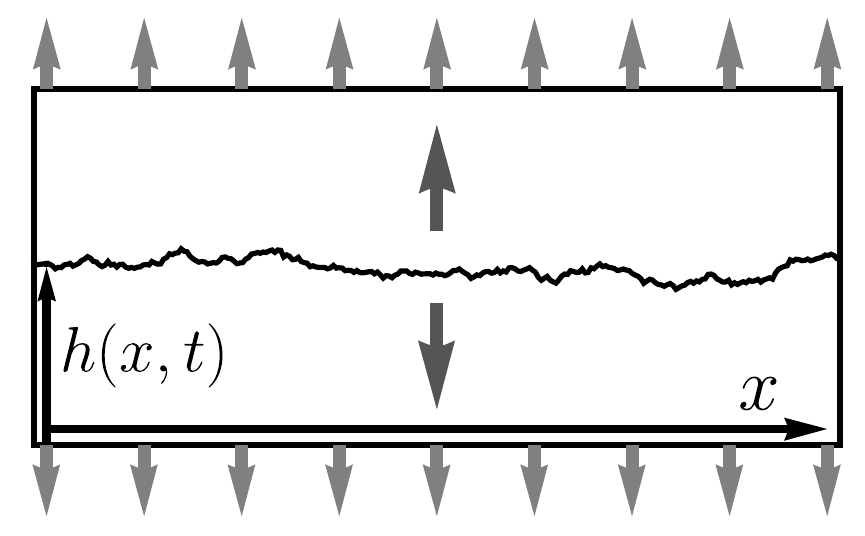}}}&{\includegraphics[width=0.45\linewidth]{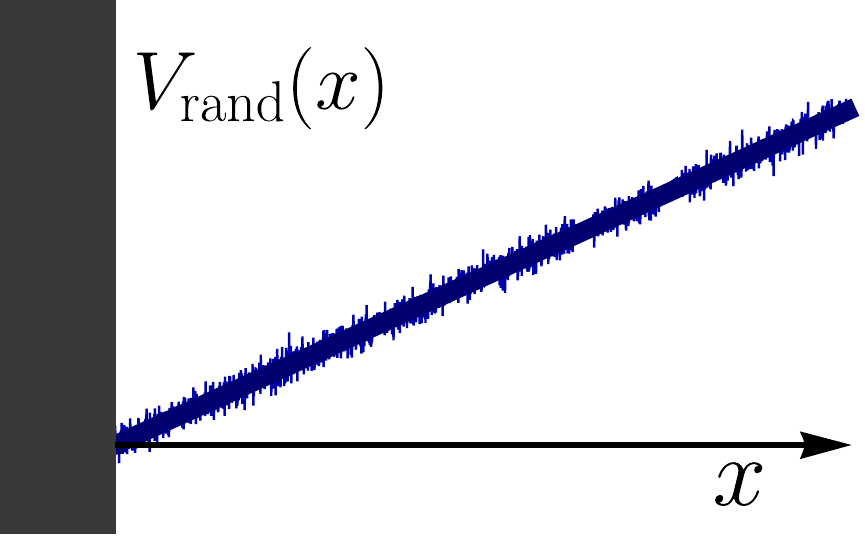}}
        \\[-2mm]
         (a) & (b) 
    \end{tabular}
\caption{Panel (a): sketch of an interface (rough black line in the middle), e.g., in a 2d Ising model prior to switching on an external field. After it is switched on, the height of the interface  $h(x,t)$ is described by the Kardar-Parisi-Zhang equation \eqref{eq:KPZ} in 1+1 dimensions. Panel (b): a realization of the random potential in the stochastic Airy Hamiltonian, see \cref{eq:Airy_op}.}
\label{fig:Tracy-widom}
\end{figure}

The first example that we consider is the \emph{Kardar--Parisi--Zhang (KPZ) equation} \cite{Kardar_1986}, which describes the stochastic dynamics of a growing interface separating a stable bulk phase from a metastable one (see Fig.~\ref{fig:Tracy-widom}(a)). In $1+1$ dimensions, the interface is parametrized by a height field $h(x,t)$, where $x \in \mathbb{R}$ denotes the spatial coordinate and $t$ the time. The KPZ equation then reads
\begin{eqnarray}\label{eq:KPZ}
\partial_t h(x,t) = \nu\, \partial_x^2 h(x,t) + \frac{\lambda_0}{2}\, (\partial_x h(x,t))^2 + \sqrt{D}\, \eta(x,t)\;.
\end{eqnarray}
Here, $\nu > 0$ is the coefficient of diffusive relaxation, $\lambda_0 > 0$ controls the strength of the nonlinear term, and $\eta(x,t)$ is a Gaussian white noise with zero mean and correlations $\langle \eta(x,t)\eta(x',t') \rangle = \delta(x-x')\delta(t-t')$, with $D>0$ {being} the noise amplitude, often interpreted as an effective temperature. When $\lambda_0 = 0$, the KPZ equation becomes linear and reduces to the well-known Edwards--Wilkinson equation\cite{Edwards_1982}. For $\lambda_0>0$, the KPZ equation is a nonlinear stochastic partial differential equation and is therefore notoriously difficult to solve, even in one dimension. Nevertheless, over the past three decades, remarkable progress in both mathematics and physics has led to a detailed understanding and to exact solutions in $1+1$ dimensions \cite{Sasamoto_2010, Dotsenko_2010, Calabrese_2010, amir2011probability,Calabrese_2011,Imamura_2012}. In particular, in the long-time limit, the height at the origin exhibits the asymptotic behavior given by
\begin{eqnarray}\label{eq:chiKPZ}
h(0,t) \simeq v_{\infty}\, t + \left(\Gamma t\right)^{1/3}\, \chi_{\rm KPZ}\;,
\end{eqnarray}
where $v_\infty$ and $\Gamma$ are deterministic constants depending on $\nu$, $\lambda_0$, and $D$, while $\chi_{\rm KPZ}$ is a time-independent random variable. While the scaling exponents governing the time dependence of $h(0,t)$~—~most notably the fluctuation exponent $1/3$~—~are universal, the coefficients $v_\infty$, $\Gamma$, and, crucially, the distribution of $\chi_{\rm KPZ}$ depend on the initial condition $h(x,0)$. As a result, the interface retains a memory of its initial state even at asymptotically long times, a striking and nontrivial signature of the nonequilibrium nature of KPZ dynamics. Different initial conditions therefore define distinct universality subclasses within the KPZ universality class. The two most common cases are:
\begin{itemize}
\item[(i)] {Flat initial condition}, $h(x,0)=0$ (see Fig.~\ref{fig:Tracy-widom}(a)): the interface remains flat on average for all $t>0$, and the random variable $\chi^{\rm flat}_{\rm KPZ}$ is exactly distributed according to the Tracy--Widom law for the GOE \cite{Calabrese_2011}, i.e.,
\begin{eqnarray} \label{chi_flat}
\chi^{\rm flat}_{\rm KPZ} = \chi_1\;.
\end{eqnarray}
\item[(ii)] {Droplet (curved) initial condition}: this leads to a curved interface profile, as in growth processes such as the Eden model \cite{Eden_1961} starting from a point seed. In this case, $\chi^{\rm drop}_{\rm KPZ}$ follows the Tracy--Widom distribution for the GUE \cite{Dotsenko_2010, Sasamoto_2010,Calabrese_2010,amir2011probability}, i.e.,
\begin{eqnarray} \label{chi_drop}
\chi^{\rm drop}_{\rm KPZ} = \chi_2\;.
\end{eqnarray}
\end{itemize}

Quite remarkably, these two universality classes~(\ref{chi_flat}) and~(\ref{chi_drop}) have been observed experimentally in a variety of physical systems, including interface growth in nematic liquid crystals~\cite{takeuchi2011growing}, coupled laser systems~\cite{fridman2012measuring}, and, more recently, dissipative many-particle systems~\cite{makey2020universality}. These observations provide striking evidence for the universality of the Tracy--Widom distributions, even though the connection between the KPZ equation~(\ref{eq:KPZ}) and the random matrix theory is far from obvious. In the following, we discuss another paradigmatic example, namely the directed polymer in a random environment. On general physical grounds, this model is believed to belong to the same KPZ universality class in $1+1$ dimensions, but in this case, as we will see, the connection to random matrix theory can be made more explicit.

The second example that we discuss here is a model of a directed polymer in a random medium -- see Fig. \ref{fig:disordered} -- introduced and solved by Johansson in Ref. \cite{Johansson_2000}. In this model, as discussed in the introductory part of these notes in \cref{sec:intro}, the energy of a configuration of a polymer starting from the origin $(0,0)$ is given by the sum of the energies $\epsilon_{i,j}$ ascribed to each site $(i,j)$ of a two-dimensional lattice that it covers (see Fig.~\ref{fig:disordered}(b))
\begin{equation}
    E(\mathcal{C}) = \sum_{\langle i,j \rangle \in \mathcal{C}}\epsilon_{i,j} \;.
\end{equation}
In this model, the energies $\epsilon_{i,j}$ are assumed to be IID random variables distributed according to an exponential distribution (as already considered above in Eq. (\ref{eq:p_exp}))
\begin{equation}
    {\rm Prob.}(\epsilon_{i,j} \leq x) = \int_{0}^x\mathrm{d}\xi \, e^{-\xi} \;.
\end{equation}
We consider here the optimal energy of the polymer, i.e., 
\begin{equation}
    E_{\mathrm{max}} = \max_{\mathcal{C}}E(\mathcal{C}) \;,
\end{equation}
where the maximum is taken over all possible polymer configurations starting at~$(0,0)$ and terminating at a {\it fixed} endpoint with coordinates $(M \geq 0,N \geq 0)$ on the square lattice (see Fig. \ref{fig:disordered}) -- this is the so-called {\it point-to-point geometry}. As  shown in Ref.~\cite{Johansson_2000}, the distribution of $E_{\max}$ coincides with the  distribution of the largest eigenvalue of a complex Laguerre-Wishart  random matrix $\mathbf{M}$, which is defined by~\cite{Wishart_1928,mehta2004random,forrester2010log}
\begin{equation}
    \mathbf{M} = \mathbf{W}\mathbf{W}^T,
    \label{eq:wish_def}
\end{equation}
where $\mathbf{W}$ is an $N \times M$ (we assume $N\leq M$) rectangular matrix with IID complex Gaussian entries.
It turns out that, similarly to Gaussian random matrices~\eqref{eq:pdf_RM_entries}, the eigenvalues ${\lambda_1, \lambda_2, \ldots, \lambda_N}$ of the Laguerre–Wishart matrices, which are real and positive as follows from Eq.~(\ref{eq:wish_def}), are statistically independent of the eigenvectors. Their joint probability distribution reads~\cite{Wishart_1928,mehta2004random,forrester2010log}
\begin{equation} \label{eq:LUE}
    P_{\mathrm{joint}}(\lambda_1, \lambda_2, \ldots, \lambda_N) \propto \prod_{i=1}^Ne^{-\lambda_i} \lambda_i^{M-N} \prod_{i<j} (\lambda_i - \lambda_j)^2 \;.
\end{equation}
The result shown in Ref.~\cite{Johansson_2000}, which connects the optimal energy of the directed polymer with the largest eigenvalue of complex Laguerre-Wishart random matrices, reads, for any finite $N$, 
\begin{equation} \label{eq:idEmin_lmax}
    E_{\mathrm{max}} \stackrel{\mbox{\footnotesize  law}}{=} \max_{1\leq i\leq N} \lambda_i,
\end{equation}
where the superscript `law' denotes an identity in law (or distribution). Hence, computing the distribution of the optimal energy of the directed polymer in the thermodynamic  limit ($N \to \infty$) amounts to computing the distribution of the largest eigenvalue of the complex Laguerre-Wishart ensemble (\ref{eq:LUE}). In fact, this model (\ref{eq:LUE}) bears a strong resemblance to the GUE in Eq.~(\ref{eq:RM_joint_evals}) with $\beta = 2$, and it can be shown that the largest eigenvalue in this ensemble, properly centered and scaled, also converges  to a Tracy-Widom random variable $\chi_2$ \cite{Johansson_2000,johnstone2001distribution}.  
The identity in Eq.~(\ref{eq:idEmin_lmax}) then implies that the cumulative distribution function of the optimal energy $E_{\max}$, properly centered and scaled, converges to the Tracy-Widom distribution ${\cal F}_2(x)$ that describes the fluctuations of the largest eigenvalue in the GUE ensemble~(\ref{eq:lambda_max}). 

So far, we have discussed applications of the Tracy--Widom distributions ${\cal F}_\beta$ associated with the classical values $\beta = 1$ and $\beta = 2$, corresponding  to the GOE and the GUE, respectively. However, as we have seen, Tracy--Widom distributions can, in fact, be defined for any value of $\beta>0$. Starting from the tridiagonal random matrix ensembles whose joint eigenvalue distribution is given by Eq.~(\ref{eq:RM_joint_evals}), one can show that the Tracy--Widom distribution describes the fluctuations of the ground-state energy of the so-called \emph{stochastic Airy operator}~\cite{Edelman_2007_stoch_Airy,Ramirez_2011_stoch_airy}
\begin{equation}
    \mathcal{H}_{\beta}  = -\frac{\mathrm{d}^2}{\mathrm{d}x^2} + x + \sqrt{\frac{2}{\beta}}\,\eta(x)\;,
    \label{eq:Airy_op}
\end{equation}
where $\eta(x)$ is a Gaussian white noise with zero mean and covariance
$\langle \eta(x)\eta(x')\rangle = \delta(x-x')$. From a physical perspective, the operator $\mathcal{H}_{\beta}$ admits a transparent interpretation as a one-dimensional Schr\"odinger operator with a random potential
\begin{eqnarray} \label{V_SAO}
V_{\rm rand}(x) = x + \sqrt{\frac{2}{\beta}}\,\eta(x)
\qquad \text{for } x>0\;,
\end{eqnarray}
with a hard-wall boundary condition at the origin, i.e., \ $V_{\rm rand}(x)=+\infty$ for $x<0$. An example of a realization of this random potential is shown in Fig.~\ref{fig:Tracy-widom}(b), corresponding to the Hamiltonian~(\ref{eq:Airy_op}). In the absence of disorder, namely in the limit $\beta\to\infty$, the noise term vanishes, and the eigenfunctions of $\mathcal{H}_{\beta\to\infty}$ reduce to the standard Airy functions, which explains the terminology `stochastic Airy operator' when finite noise/disorder is present. This thus provides a natural and physically meaningful realization of Tracy--Widom distributions for arbitrary $\beta>0$, with direct connections to localization phenomena in one-dimensional quantum disordered systems.

\section{Conclusions and outlook}
\label{sec:conclusions}

These notes have presented a brief introduction to extreme value statistics from the perspective of statistical physics, with an emphasis on disordered systems and correlated stochastic processes. A central objective has been to show how problems involving extrema of time series can be reformulated in terms of quantities from statistical mechanics, such as partition functions of interacting particle systems or survival probabilities of stochastic processes. This reformulation provides a concrete framework in which extreme value statistics can be analyzed using tools and concepts familiar from statistical physics.

Several applications of these ideas have been discussed explicitly throughout the notes, illustrating how extreme value statistics naturally arise in specific models of disordered systems and nonequilibrium stochastic dynamics. These examples show how correlations, interactions, and dynamical constraints can strongly affect the statistics of extremes and lead to behaviors that differ markedly from the classical independent-variable setting.

Beyond the cases treated here, a number of related extensions can be considered. For instance, stochastic processes with resetting \cite{Evans_2011_PRL,Evans_2020,pal2022inspection, gupta2022stochastic}  in which the dynamics is intermittently restarted from a prescribed state, offer a simple and physically motivated modification that can significantly alter extreme value statistics, while remaining analytically accessible \cite{SP21,GYC23,HMS24,DHM26}. Another class of processes that have recently attracted some attention involve conditionally independent and identically distributed (CIID) variables~\cite{Majumdar_2026_arxiv}.  They provide another class of models, interpolating between independent and strongly correlated systems, that allow for explicit calculations in physically relevant settings.
 More generally, we hope that the approaches presented in these notes will be useful for a broader range of systems in statistical physics where extreme events play a role. 

\vspace*{0.5cm}
\noindent{\bf Acknowledgments:} We would like to thank the organizers of the school FPSP XVI: A. Gambassi, V. Ros and T. Schilling. GS acknowledges support from ANR Grant No.~ANR-23-CE30-0020-01 EDIPS.    

\section*{References}
\addtocontents{toc}{\protect\fixappendix}

\providecommand{\newblock}{}

\end{document}